\documentclass[10pt,USenglish]{article}
\pdfoutput=1

\usepackage{cite}

\renewcommand{\Phi}{\phi}

\newcommand{\diagentry}[1]{\mathmakebox[3em]{#1}}

\usepackage[latin9]{inputenc}
\usepackage{geometry}
\usepackage[USenglish]{babel}
\usepackage{verbatim}
\usepackage{prettyref}
\usepackage{amsmath,mathtools}
\usepackage{mathdots}
\usepackage[normalem]{ulem}
\usepackage{amssymb}
\usepackage{graphicx}
\usepackage{setspace}
\usepackage{esint}
\usepackage{bbold}
\usepackage{ dsfont }
\usepackage{pdfpages}

\usepackage{color}
\definecolor{darkgreen}{rgb}{0,0.5,0}
\definecolor{darkblue}{rgb}{0,0,0.6}
\definecolor{purple}{rgb}{0.4,.2,0.7}
\definecolor{awesome}{rgb}{1.0, 0.13, 0.32}

\usepackage[colorlinks=true,citecolor=darkgreen,linkcolor=purple,urlcolor=purple]{hyperref}

\numberwithin{equation}{section}

\begin{document}
\begin{flushright}
{\tt MIT-CTP-4877}
\end{flushright}

~
\vskip15mm

\begin{center} {\Large \textsc{Mini-BFSS in Silico}}

\vskip15mm

Tarek Anous$^{1,2}$ and Cameron Cogburn$^3$

\vskip5mm


\it{$^1$Department of Physics and Astronomy, University of British Columbia, 6224 Agricultural Road, Vancouver, B.C. V6T 1Z1, Canada}\\
\it{$^2$Center for Theoretical Physics, Massachusetts Institute of Technology, Cambridge, MA 02139, USA}\\
\it{$^3$Department of Physics, Massachusetts Institute of Technology, Cambridge, MA 02139, USA}

\vskip5mm

\tt{ tarek@phas.ubc.ca, ccogburn@mit.edu}

\end{center}

\vskip15mm

\begin{abstract}
We study a mass-deformed $\mathcal{N}=4$ version of the BFSS matrix model with three matrices and gauge group $SU(2)$. This model has zero Witten index. Despite this, we give numerical evidence for the existence of four supersymmetric ground states, two bosonic and two fermionic, in the limit where the mass deformation is tuned to zero.

\end{abstract}


\pagebreak
\pagestyle{plain}

\setcounter{tocdepth}{2}

\tableofcontents

\section{Introduction}
This paper concerns itself with the supersymmetric quantum mechanics of three bosonic $SU(N)$ matrices and their fermionic superpartners. The model in question, introduced in \cite{Claudson:1984th,Denef:2002ru,Asplund:2015yda}, has four supercharges and describes the low energy effective dynamics of a stack of $N$ wrapped D-branes in a string compactification down to 3+1 dimensions. When the compactification manifold has curvature and carries magnetic fluxes, the bosonic matrices obtain masses\cite{Asplund:2015yda}. When the compact manifold is Calabi-Yau and carries no fluxes, the matrices are massless. 

This theory has flat directions whenever the matrices are massless, and hence is a simplified version of the BFSS matrix model \cite{Banks:1996vh}, which, for the sake of comparison, has nine bosonic $SU(N)$ matrices and 16 supercharges and describes the non-Abelian geometry felt by D-particles in a non-compact 9+1 dimensional spacetime. We hence dub the model studied here: mini-BFSS (or mini-BMN \cite{Berenstein:2002jq} in the massive case). The Witten index $W_I$ has been computed for mini-BFSS  \cite{Sethi:1997pa,Moore:1998et,Yi:1997eg,Lee:2016dbm,deWit:1997ab} and vanishes, meaning that the existence of supersymmetric ground states is still an open question. Even the refined index, twisted by a combination of global symmetries and calculated in \cite{Lee:2016dbm}, gives us little information about the set of ground states due to the subtleties associated with computing such indices in the presence of flat directions in the potential. This is in stark contrast with the full BFSS model, whose Witten index $W_I=1$, implying beyond doubt the existence of at least one supersymmetric ground state. The zero index result for mini-BFSS has led to the interpretation that it may not have any zero energy ground states \cite{Sethi:1997pa,deWit:1997ab}, and hence no holographic interpretation. The logic being that, without a rich low energy spectrum, scattering in mini-BFSS would not mimic supergraviton scattering in a putative supersymmetric holographic dual \cite{Polchinski:1999br}.
Of course 
a vanishing $W_I$ does not confirm the absence of supersymmetric ground states---as there may potentially be an exact degeneracy between the bosonic and fermionic states at zero energy.

We weigh in on the existence of supersymmetric states in mini-BFSS by solving the Schr\"{o}dinger equation numerically for the low-lying spectrum of the $N=2$ model, in the \emph{in silico} spirit of \cite{Anous:2015xah}. To deal with the flat directions we numerically diagonalize the Hamiltonian of the mass-deformed mini-BMN matrix model, for which the flat directions are absent, and study the bound state energies as a function of the mass. A numerical analysis of mini-BFSS can also be found in \cite{Wosiek:2002nm,Campostrini:2004bs} which use different methods.

What we uncover is quite surprising. As we tune the mass parameter $m$ to zero, we find evidence for four supersymmetric ground states, two bosonic and two fermionic, which cancel in the evaluation of $W_I$. This result seems to agree with plots found in \cite{Wosiek:2002nm,Campostrini:2004bs}. It must be said that our result does not constitute an existence proof for supersymmetric threshold bound states in the massless limit, but certainly motivates a further study of the low-lying spectrum of these theories. 

The organization of the paper is as follows: in section \ref{sec:setup} we present the supercharges, Hamiltonian and symmetry generators of the mini-BMN model for arbitrary $N$. In section \ref{sec:su2} we restrict to $N=2$ and give coordinates in which the Schr\"{o}dinger equation becomes separable. In section \ref{sec:numerics} we provide our numerical results and in section \ref{sec:effectiv} we derive the one-loop effective theory on the moduli space in the massless theory. We conclude with implications for the large-$N$ mini-BFSS model in section \ref{sec:discussion}. We collect formulae for the Schr\"{o}dinger operators maximally reduced via symmetries in appendix \ref{ap:schroeq} and compute the one-loop metric on the Coulomb branch moduli space in appendix \ref{ap:metric}.

\section{Setup}\label{sec:setup}
\subsection{Supercharges and Hamiltonian}
Let us consider a supersymmetric quantum mechanics of $SU(N)$ bosonic matrices $X^i_A$ and their superpartners $\lambda_{A\alpha}$. The quantum mechanics we have in mind has four supercharges:\footnote{Spinors and their conjugates transform respectively in the $\boldsymbol{2}$ and  $\bar{\boldsymbol{2}}$ of of $SO(3)$. Spinor indices are raised and lowered using the Levi-Civita symbol $\epsilon^{\alpha\beta}=-\epsilon_{\alpha\beta}$ with $\epsilon^{12}=1$. Thus in our conventions:
\begin{equation}
\left(\bar{\psi}\epsilon\right)_\alpha=\bar{\psi}^{\gamma}\epsilon_{\gamma\alpha}~,\quad\quad\quad\quad \left(\epsilon\psi\right)^\alpha=\epsilon^{\alpha\gamma}\psi_\gamma~,\quad\quad\quad\quad \epsilon_{\alpha\omega}\epsilon^{\omega\beta}=\delta_\alpha^{~\,\beta}~.
\end{equation}}
\begin{equation}
Q_{\alpha}=\left(-i\partial_{X^i_A}-i\,m\,X^{i}_A-i\,W^i_A\right)\boldsymbol{\sigma}^{i~\gamma}_{\alpha}\lambda_{A\gamma}~,\quad\quad\bar{Q}^\beta=\bar{\lambda}^{~\gamma}_A\boldsymbol{\sigma}^{i~\beta}_\gamma\left(-i\partial_{X^i_A}+i\,m\,X^i_A+i\,W^i_A\right)~.
\end{equation}
The parameter $m$ is simply the mass of $X^i_A$. The massless version of this model was introduced in \cite{Claudson:1984th} and can be derived by dimensionally reducing $\mathcal{N}=1,~d=4$ super Yang-Mills to the quantum mechanics of its zero-modes. The mass deformation was introduced in \cite{Asplund:2015yda}, and can be obtained from a dimensional reduction of the same gauge theory on $\mathds{R}\times S^3$. We direct the reader to \cite{Denef:2002ru,Asplund:2015yda} for an introduction to these models. This quantum mechanics should be thought of as a simplified version of the BMN matrix model \cite{Berenstein:2002jq} (mini-BMN for brevity). The massless limit should then be thought of as a mini-BFSS matrix model \cite{Banks:1996vh}. The lowercase index $i=1,\dots,3$ runs over the spatial dimensions (in the language of the original gauge theory), and the uppercase index $A=1,\dots,N^2-1$, runs over the generators of the gauge group $SU(N)$.  The $\boldsymbol{\sigma}^i$ are the Pauli matrices and greek indices run over $\alpha=1,2$. In keeping with \cite{Claudson:1984th}, we have defined $W^i_A\equiv\partial W/\partial X^i_A$ where
\begin{equation}
W\equiv\frac{g}{6}f_{ABC}\,\epsilon_{ijk}\,X^i_A\,X^j_B\,X^k_C,
\end{equation}
and $f_{ABC}$ are the structure constants of $SU(N)$. 
The gauginos obey the canonical fermionic commutation relations $\left\{\lambda_{A\alpha},\bar{\lambda}^\beta_B\right\}=\delta_{AB}\delta_\alpha^{~\beta}$, and hence the algebra generated by these supercharges is \cite{Asplund:2015yda}
\begin{equation}\label{susyalg}
\left\{Q_\alpha,\bar{Q}^\beta\right\}=2\left(\delta_\alpha^{~\beta}\,H-g\,\boldsymbol{\sigma}^{k~\beta}_\alpha\,X^k_A\,G_A+\,m\,\boldsymbol{\sigma}^{k~\beta}_\alpha J^k\right)~,\quad\quad\{\bar{Q}^\alpha,\bar{Q}^\beta\}=\left\{Q_\alpha,Q_\beta\right\}=0~,
\end{equation}
with Hamiltonian:
\begin{equation}\label{eq:fullhamiltonian}
H\equiv -\frac{1}{2}\partial_{X^i_A}\partial_{X^i_A}+\frac{1}{2}m^2 \left(X^i_A\right)^2+m\,X^i_A\,W^i_A+\frac{g^2}{4}\left(f_{ABC}\,X^i_B\,X^j_C\right)^2-\frac{3}{4}m[\bar{\lambda}_A,\lambda_{A}]+ig\,f_{ABC}\bar{\lambda}_A\,X^k_B\,\boldsymbol{\sigma}^k\lambda_C~.
\end{equation}
The operators $G_A$ and $J^k$ appearing in the algebra are, respectively, the generators of gauge transformations and $SO(3)$ rotations. These are given by:
\begin{equation}\label{gaugerot}
G_A\equiv -i\,f_{ABC}\left(X^i_B\,\partial_{X^i_C}+\bar{\lambda}_B\lambda_C\right)~, \quad\quad J^i\equiv-i\,\epsilon_{ijk}\,X^j_A\,\partial_{X^k_A}+\frac{1}{2}\bar{\lambda}_A\boldsymbol{\sigma}^i\lambda_A~.
\end{equation}
In solving for the spectrum of this theory, we must impose the constraint $G_A|\psi\rangle=0~, \forall~A$. In the above expressions, whenever fermionic indices are suppressed, it implies that they are being summed over.

Let us briefly note the dimensions of the fields and parameters in units of the energy $[\mathcal{E}]=1$. These are $[X]=-1/2$, $[\lambda]=0$, $[g]=3/2$ and $[m]=1$. Therefore, an important role will be played by the dimensionless quantity
\begin{equation}\label{eq:nu}
\nu\equiv \frac{m}{g^{2/3}}~.
\end{equation}

We consider here the mass deformed gauge quantum mechanics because, in the absence of the mass parameter $m$, the classical potential has flat directions (see figure \ref{fig:constpot}). Turning on this mass deformation gives us a dimensionless parameter $\nu$, to tune in studying the spectrum of this theory, and allows us to approach the massless limit from above. 

\subsection{Symmetry algebra}

Let us now give the symmetry algebra of the theory. The components of $\vec{{J}}$ satisfy:
\begin{align}
&\left[J^i,J^j\right]=i\,\epsilon_{ijk}\,{J}^k~, &&\left[{J}^i,Q_\alpha\right]=-\frac{1}{2}\boldsymbol{\sigma}^{i~\gamma}_\alpha Q_\gamma~,\nonumber\\
&\left[\vec{{J}}^{\,2},{J}^i\right]=0~, &&\left[{J}^i,\bar{Q}^\alpha\right]=\frac{1}{2}\bar{Q}^\beta\,\boldsymbol{\sigma}^{i~\alpha}_\beta ~.\label{jalg}
\end{align}
There is an additional $U(1)_R$ generator $R\equiv\bar{\lambda}_A\lambda_A$ which counts the number of fermions. It satisfies
\begin{equation}
[R,Q_\alpha]=-Q_\alpha~,\quad\quad\quad\quad\left[R,\bar{Q}^\alpha\right]=+\bar{Q}^\alpha~,\quad\quad\quad\quad\left[R,{J}^i\right]=0~.\label{ralg}
\end{equation}
The Hamiltonian also has a particle-hole symmetry:
\begin{equation}\label{discretesym}
\bar{\lambda}^\alpha_A\rightarrow \epsilon^{\alpha\gamma}\lambda_{A\gamma}~,\quad\quad \lambda_{A\alpha}\rightarrow \bar{\lambda}_A^\gamma\epsilon_{\gamma\alpha}~,\quad\quad \epsilon^{12}=-\epsilon_{12}=1~,
\end{equation}
where $\epsilon^{\alpha\beta}$ is the Levi-Civita symbol. This transformation leaves the Hamiltonian invariant but takes $R\rightarrow 2(N^2-1)-R$ and effectively cuts our problem in half.

One peculiar feature of the mass deformed theory is that the supercharges do not commute with the Hamiltonian as a result of the vector $\vec{J}$ appearing in (\ref{susyalg}). It is easy to show that
\begin{equation}
\big[H,Q_\alpha\big]=\frac{m}{2}Q_\alpha~,\quad\quad\quad \big[H,\bar{Q}^\beta\big]=-\frac{m}{2}\bar{Q}^\beta~.
\end{equation}
Thus, acting with a supercharge increases/decreases the energy of a state by $\pm \frac{m}{2}$. This is a question of $R$-frames, as discussed in \cite{Asplund:2015yda}. Essentially we can choose to measure energies with respect to the shifted Hamiltonian $H_m\equiv H+\frac{m}{2}R$, which commutes with the supercharges, and write the algebra as:
\begin{equation}
\left\{Q_\alpha,\bar{Q}^\beta\right\}=2\left\lbrace\delta_\alpha^{~\beta}\,\left(H_m-\frac{m}{2}R\right)-g\,\boldsymbol{\sigma}^{k~\beta}_\alpha\,X^k_A\,G_A+\,m\,\boldsymbol{\sigma}^{k~\beta}_\alpha J^k\right\rbrace~.
\end{equation}


\subsection{Interpretation as D-particles}
The $\nu\rightarrow 0$ limit of this model can be thought of as the worldvolume theory of a stack of $N$ D-branes compactified along a special Lagrangian cycle of a Calabi-Yau three-fold \cite{Denef:2002ru}. The $X^i_A$ then parametrize the non-Abelian geometry felt by the compactified D-particles in the remaining non-compact $3+1$ dimensional asymptotically flat spacetime. The addition of the mass parameter corresponds to adding curvature and magnetic fluxes to the compact manifold \cite{Asplund:2015yda} 
changing the asymptotics of the non-compact spacetime to AdS$_4$. This interpretation was argued in \cite{Anninos:2013mfa,Asplund:2015yda} and passes several consistency checks. Hence we should think of the mass deformed theory as describing the non-relativistic dynamics of D-particles in an asymptotically AdS$_4$ spacetime and the massless limit as taking the AdS radius to infinity in units of the string length.

To be more specific, it will be useful to translate between our conventions and the conventions of \cite{Asplund:2015yda}. One identifies $m=\Omega$, $g^2=1/m_v$, $\{X,\lambda\}_{\rm us}= m_v^{1/2}\{X,\lambda\}_{\rm them}$ in units where the string length $l_s=1$. Reintroducing $l_s$, this dictionary implies that $g^2=g_s /l_s^3\sqrt{2\pi}$, with $g_s$ the string coupling, gets set by a combination of the magnetic fluxes threading the compact manifold and similarly $\ell_{\rm AdS}\equiv 1/m$ gets set by a combination of these magnetic fluxes and the string length. For AdS$_4\times\mathds{C}P^3$ compactifications dual to ABJM this was worked out in detail in \cite{Asplund:2015yda} and they identify
\begin{equation}
g_s=\left(\frac{32\pi^2N}{k^5}\right)^{\frac{1}{4}}~,\quad\quad\ell_{\rm AdS}=\left(\frac{N}{8\pi^2 k}\right)^{\frac{1}{4}}l_s~,
\end{equation}
where $k$ and $N$ are, respectively, integrally quantized magnetic 2-form and 6-form flux. In this example taking $\nu=\sqrt{2\pi}\left(k^2/N\right)^{1/3}\rightarrow 0$ while keeping $g_s$ fixed takes the AdS radius to infinity in units of $l_s$.

The main focus of the next sections is on whether this stack of D-particles forms a supersymmetric bound state, particularly in the $\nu\rightarrow 0$ limit. There the Witten index $W_I\equiv \text{Tr}_\mathcal{H}\left\lbrace (-1)^R\,e^{-\beta\,H}\right\rbrace$ has been computed \cite{Sethi:1997pa,Moore:1998et,Yi:1997eg,Lee:2016dbm,deWit:1997ab} and evaluates to zero. This is in contrast with the full BFSS matrix model, whose index is $W_I=1$, confirming the existence of a supersymmetric ground state. We will use the numerical approach of \cite{Anous:2015xah} and verify if supersymmetry is preserved or broken in the $SU(2)$ case. We find evidence that supersymmetry is preserved in the $\nu\rightarrow 0$ limit, and that there are precisely 4 ground states contributing to the vanishing Witten index.

\section{Quantizing the $SU(2)$ theory}\label{sec:su2}

\subsection{Polar representation of the matrices}

We are aiming to solve the Schr\"{o}dinger problem $H_m|\psi\rangle=\mathcal{E}_m|\psi\rangle$. We will not be able to do this for arbitrary $N$ and from here on we will restrict to gauge group $SU(2)$ for which the structure constants $f_{ABC}=\epsilon_{ABC}$. In this case the wavefunctions depend on 9 bosonic degrees of freedom tensored into a 64-dimensional fermionic Hilbert space.  It is thus incumbent upon us to reduce this problem maximally via symmetry. In order to do so, we exploit the fact that the matrices $X^i_A$ admit a polar decomposition as follows
\begin{equation}\label{polardecomp}
X_A^i=L_{AB}\, \Lambda_{B}^j\, M^{\text{T}\,ji}
\end{equation}
with
\begin{equation}
L\equiv e^{-i \varphi_1\, \boldsymbol{\mathcal{L}}^3}e^{-i \varphi_2\, \boldsymbol{\mathcal{L}}^2}e^{-i \varphi_3\, \boldsymbol{\mathcal{L}}^3}~,\quad\quad\quad M\equiv e^{-i \vartheta_1\, \boldsymbol{\mathcal{L}}^3}e^{-i \vartheta_2\, \boldsymbol{\mathcal{L}}^2}e^{-i \vartheta_3\, \boldsymbol{\mathcal{L}}^3}~,
\end{equation}
and  $\left[\boldsymbol{\mathcal{L}}^i\right]_{jk}\equiv-i\epsilon_{ijk}$ are the generators of $SO(3)$. The diagonal matrix
\begin{equation}
\Lambda\equiv\text{diag}(\mathbf{x}_1,\mathbf{x}_2,\mathbf{x}_3)
\end{equation}
represents the spatial separation between the pair of D-branes in the stack. The $\varphi_i$ and $\vartheta_i$ represent the (respectively gauge-dependent and gauge-independent) Euler-angle rigid body rotations of the configuration space. This parametrization is useful because the Schr\"{o}dinger equation is separable in these variables, as we show in appendix \ref{ap:schroeq}.

The metric on configuration space can be re-expressed as:
\begin{align}
\sum_{A,i}dX_A^i\,dX_A^i&=\sum_{a=1}^3 d\mathbf{x}^2_a+ I_a\left(d\Omega_a^2+d\omega_a^2\right)-2K_a\,d\Omega_a\,d\omega_a~,\\
I_a&\equiv \mathbf{x}_b\,\mathbf{x}_b-\mathbf{x}_a^2~,\quad\quad\quad K_a\equiv|\epsilon_{abc}|\,\mathbf{x}_b\,\mathbf{x}_c~.
\end{align}
The angular differentials are the usual $SU(2)$ Cartan-Maurer differential forms defined as follows:
\begin{equation}
d\omega_a=-\frac{1}{2}\epsilon_{abc}\,\left[L^T\cdot dL\right]_{bc}~,\quad\quad\quad d\Omega_a=-\frac{1}{2}\epsilon_{abc}\left[M^T\cdot dM\right]_{bc}~.
\end{equation}
The volume element used to compute the norm of the wavefunction is
\begin{equation}
\prod_{i,A} dX_{A}^idX_A^i=\Delta(\mathbf{x}_a)\prod_{i=1}^3 d\mathbf{x}_i\sin\varphi_2\prod_{j=1}^3 d\varphi_j\,\sin\vartheta_2\prod_{k=1}^3 d\vartheta_k \,,
\end{equation}
where $\Delta(\mathbf{x}_a)\equiv\left(\mathbf{x}_1^2-\mathbf{x}_2^2\right)\left(\mathbf{x}_3^2-\mathbf{x}_2^2\right)\left(\mathbf{x}_3^2-\mathbf{x}_1^2\right)$ is the Vandermonde determinant with squared eigenvalues. To cover the configuration space correctly, we take the new coordinates to lie in the range \cite{Smilga:1984jg}:
\begin{equation}
\mathbf{x}_3\geq\mathbf{x}_1\geq|\mathbf{x}_2|\geq0~,\quad\quad \pi\geq\varphi_2\,,\vartheta_2\geq0~,\quad\quad 2\pi\geq \varphi_{i\neq2}\,,\vartheta_{i\neq2}\geq0~.
\end{equation}
The generators of gauge-transformations $G_A$ and rotations  $J^i$ are given in (\ref{gaugerot}). These satisfy
\begin{align}
&\left[J^i,J^j\right]=i\,\epsilon_{ijk}\,{J}^k~, &&\left[G_A,G_B\right]=i\,\epsilon_{ABC}\,G_C~,\nonumber\\
&\left[\vec{{J}}^{\,2},{J}^i\right]=0~, &&\left[{J}^i,G_A\right]=0 ~.
\end{align}
To label the $SU(2)_{\text{gauge}}\times SO(3)_J$ representations of the wavefunctions, it is useful to define the ``body fixed" angular momentum and gauge operators $\vec{P}\equiv M^{-1}\cdot \vec{J}$ and $\vec{S}\equiv L^{-1}\cdot \vec{G}$, which satisfy
\begin{align}
&\vec{P}^2=\vec{J}^2~, &&\vec{S}^2=\vec{G}^2~,\nonumber\\
&\left[P^i,P^j\right]=-i\,\epsilon_{ijk}\,{P}^k~, &&\left[S_A,S_B\right]=-i\,\epsilon_{ABC}\,S_C~,\nonumber\\
&\left[P^i,J^j\right]=0~, &&\left[S_A,G_B\right]=0~.
\end{align}
Unlike the generators of angular momentum, $\vec{P}$ is not conserved. However, as we explain in appendix \ref{ap:schroeq}, it is still useful for separating variables.

 Let us give expressions for the bosonic parts of $\vec{J}$ and $\vec{P}$, which we call $\vec{\mathcal{J}}$ and $\vec{\mathcal{P}}$ respectively, in terms of the angular coordinates. These are:
\begin{align}
\mathcal{J}^1&=-i\left(-\cos\vartheta_1\cot\vartheta_2\,\partial_{\vartheta_1}-\sin\vartheta_1\partial_{\vartheta_2}+\frac{\cos\vartheta_1}{\sin\vartheta_2}\,\partial_{\vartheta_3}\right)~,\\
\mathcal{J}^2&=-i\left(-\sin\vartheta_1\cot\vartheta_2\,\partial_{\vartheta_1}+\cos\vartheta_1\partial_{\vartheta_2}+\frac{\sin\vartheta_1}{\sin\vartheta_2}\,\partial_{\vartheta_3}\right)~,\\
\mathcal{J}^3&=-i\,\partial_{\vartheta_1}~,
\end{align}
and
\begin{align}
\mathcal{P}^1&=-i\left(-\frac{\cos\vartheta_3}{\sin\vartheta_2}\,\partial_{\vartheta_1}+\sin\vartheta_3\partial_{\vartheta_2}+\cot\vartheta_2\,\cos\vartheta_3\,\partial_{\vartheta_3}\right)~,\\
\mathcal{P}^2&=-i\left(~~\frac{\sin\vartheta_3}{\sin\vartheta_2}\,\partial_{\vartheta_1}+\cos\vartheta_3\partial_{\vartheta_2}-\cot\vartheta_2\,\sin\vartheta_3\,\partial_{\vartheta_3}\right)~,\\
\mathcal{P}^3&=-i\,\partial_{\vartheta_3}~.
\end{align}
Similarly let us define $\mathcal{G}_A$ and $\mathcal{S}_A$ as the bosonic parts of the the $G_A$ and $S_A$ operators. The $\mathcal{G}_A$ are related to the $\mathcal{J}^i$ by replacing $\vartheta_i\rightarrow\varphi_i$. It is easy to guess that the $\mathcal{S}_A$ are then related to the $\mathcal{P}^i$ via the same replacement.

\begin{figure}
  \begin{centering}
  \includegraphics[width=0.42\textwidth]{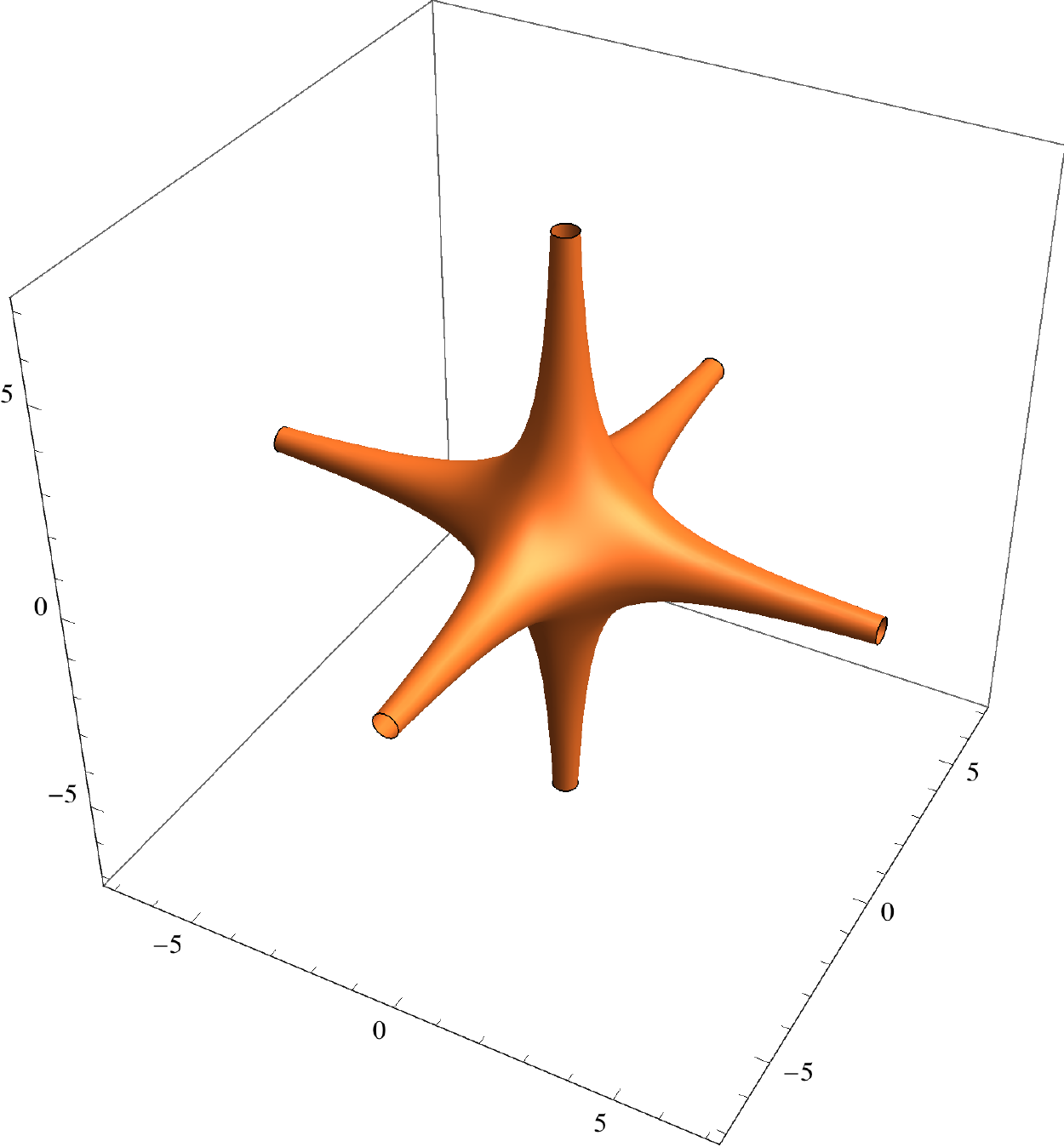}\hfill
  \includegraphics[width=0.42\textwidth]{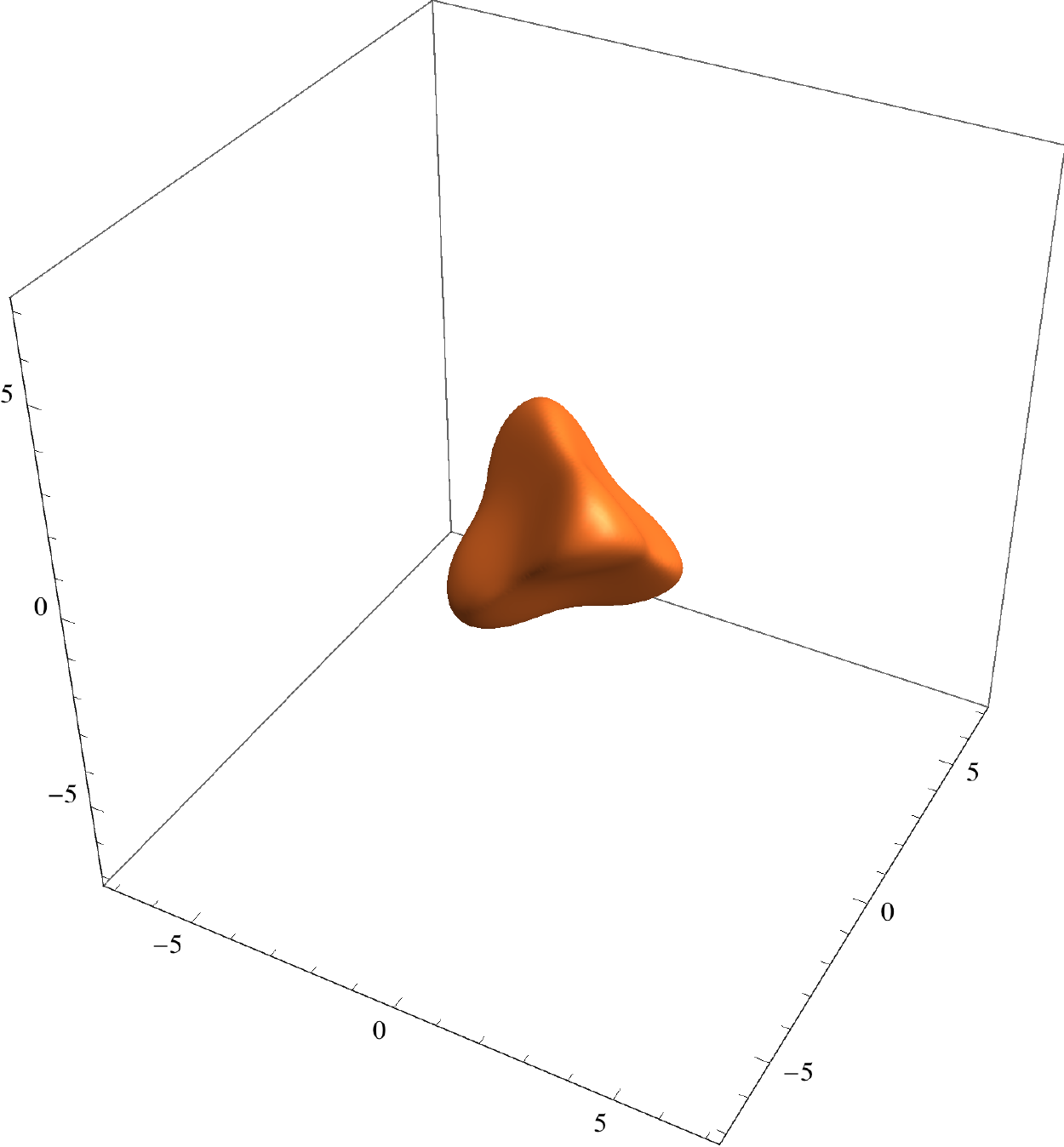}
  \caption{Contours of constant potential energy $V=2$ in units where $g=1$ as a function of $\mathbf{x}_a$. The left hand figure is evaluated at $m=0$ whereas the right hand figure is evaluated at $m=1$. The long spikes in the left figure are indicative of the flat directions along the moduli space. These flat directions get lifted for any finite $m$. }\label{fig:constpot}
\end{centering}
\end{figure}

We are now ready to give expressions for the momentum operators and the kinetic energy operator in terms of the new variables. These are \cite{Goldstone:1978he}:
\begin{align}
-i\partial_{X^i_A}&=-iL_{Aa}M^{ib}\left\lbrace\delta_{ab}\,\partial_{\mathbf{x}_a}+i\frac{\epsilon_{abc}}{\mathbf{x}_a^2-\mathbf{x}_b^2}\left(\mathbf{x}_a\,\mathcal{P}^c+\mathbf{x}_b\,\mathcal{S}_c\right)\right\rbrace~,\\
-\frac{1}{2}\partial_{X^i_A}\partial_{X^i_A}&=-\frac{1}{2\Delta}\partial_{\mathbf{x}_a}\Delta\partial_{\mathbf{x}_a}+\frac{1}{2}\sum_{a=1}^3\frac{I_a(\mathcal{P}^{a2}+\mathcal{S}_a^{2})+2K_a \mathcal{P}^a\mathcal{S}_a}{I_a^2-K_a^2}~.\label{kin}
\end{align}
It is also straightforward to write down the bosonic potential $V$ in terms of the new variables:
\begin{equation}
V=\frac{1}{2}m^2\,\mathbf{x}_a\,\mathbf{x}_a+3g\,m\,\mathbf{x}_1\mathbf{x}_2\mathbf{x}_3+\frac{g^2}{2}\left(\mathbf{x}_1^2\mathbf{x}_2^2+\mathbf{x}_1^2\mathbf{x}_3^2+\mathbf{x}_2^2\mathbf{x}_3^2\right)~.\label{pot}
\end{equation}
As expected it is independent of the angular variables. We have depicted constant potential surfaces in figure \ref{fig:constpot}.

Apart from the coordinates $\mathbf{x}_a$ the following non-linear coordinates will often appear in the equations below:
\begin{equation}\label{nonlindefs}
\mathbf{y}_a\equiv\frac{I_a}{I_a^2-K_a^2}=\frac{1}{2}|\epsilon_{abc}|\frac{\mathbf{x}_b^2+\mathbf{x}_c^2}{\left(\mathbf{x}_b^2-\mathbf{x}_c^2\right)^2}~,\quad\quad\quad\quad\mathbf{z}_a\equiv\frac{K_a}{I_a^2-K_a^2}=|\epsilon_{abc}|\frac{\mathbf{x}_b\,\mathbf{x}_c}{\left(\mathbf{x}_b^2-\mathbf{x}_c^2\right)^2}~.
\end{equation}
With these definitions the kinetic term can be written as:
\begin{equation}
-\frac{1}{2}\partial_{X^i_A}\partial_{X^i_A}=-\frac{1}{2\Delta}\partial_{\mathbf{x}_a}\Delta\partial_{\mathbf{x}_a}+\frac{1}{2}\left[\mathbf{y}_a(\mathcal{P}^{a2}+\mathcal{S}_a^{2})+2\,\mathbf{z}_a\, \mathcal{P}^a\mathcal{S}_a\right]~.\nonumber
\end{equation}
Notice that the term $\sum_{a=1}^3\mathbf{y}_a\,\mathcal{P}^{a2}$ is the kinetic energy of a rigid rotor with principal moments of inertial $\mathbf{y}_a^{-1}$.
Unlike the $c=1$ matrix model, the angular-independent piece of the kinetic term can not be trivialized by absorbing a factor of $\sqrt{\Delta}$ into the wavefunction \cite{Klebanov:1991qa}. Instead we have:
\begin{equation}
  -\frac{1}{2\Delta}\partial_{\mathbf{x}_a}\Delta\partial_{\mathbf{x}_a}=-\frac{1}{2}\left(\frac{1}{\sqrt{\Delta}}\partial_{\mathbf{x}_a}^2\sqrt{\Delta}+T\right)~,
\end{equation}
where
\begin{equation}\label{eq:Tdef}
T\equiv\sum_{a=1}^3\mathbf{y}_a=\frac{\mathbf{x}_1^2+\mathbf{x}_2^2}{\left(\mathbf{x}_1^2-\mathbf{x}_2^2\right)^2}+\frac{\mathbf{x}_1^2+\mathbf{x}_3^2}{\left(\mathbf{x}_1^2-\mathbf{x}_3^2\right)^2}+\frac{\mathbf{x}_2^2+\mathbf{x}_3^2}{\left(\mathbf{x}_3^2-\mathbf{x}_3^2\right)^2}~,
\end{equation}
and its appearance in the Schr\"{o}dinger equation acts as an attractive effective potential between the $\mathbf{x}_a$.

\subsection{Gauge-invariant fermions}
Because the operators $G_A$ in (\ref{gaugerot}) have a nontrivial dependence on the gauginos $\lambda_{A\alpha}$ it is not sufficient to suppress the wavefunction's dependence on gauge angles $\varphi_i$ entirely. Instead we can write down a set of gauge-invariant fermions that will contain the entire dependence on the gauge angles \cite{Halpern:1997fv}:
\begin{equation}\label{gaugeinvferm}
\chi_{A\alpha}\equiv L_{BA}\lambda_{B\alpha}~,\quad\quad\quad\bar{\chi}_A^\beta\equiv L_{BA}\bar{\lambda}_{B}^\beta~.
\end{equation}
These satisfy $\left\{\chi_{A\alpha},\bar{\chi}_B^\beta\right\}=\delta_{AB}\delta_\alpha^{~\beta}$, but no longer commute with bosonic derivatives.
Defining  $\tilde{\boldsymbol{\sigma}}^{i~\beta}_\alpha\equiv M^{ji}\boldsymbol{\sigma}^{j~\beta}_\alpha$, we can now write the supercharges in terms of the new parametrization. These are:
\begin{align}
Q_\alpha=-i\,\tilde{\boldsymbol{\sigma}}_\alpha^{b~\gamma}\chi_{a\gamma}\left(\delta_{ab}\left\lbrace\partial_{\mathbf{x}_b}+m\,\mathbf{x}_b+\frac{g}{2}|\epsilon_{bst}|\,\mathbf{x}_s\mathbf{x}_t\right\rbrace+i\frac{\epsilon_{abc}}{\mathbf{x}_a^2-\mathbf{x}_b^2}\left(\mathbf{x}_a\,\mathcal{P}^c+\mathbf{x}_b\,\mathcal{S}_c\right)\right)~,\nonumber\\
\bar{Q}^\beta=-i\,\bar{\chi}_{a}^{\gamma}\,\tilde{\boldsymbol{\sigma}}_\gamma^{b~\beta}\left(\delta_{ab}\left\lbrace\partial_{\mathbf{x}_b}-m\,\mathbf{x}_b-\frac{g}{2}|\epsilon_{bst}|\,\mathbf{x}_s\mathbf{x}_t\right\rbrace+i\frac{\epsilon_{abc}}{\mathbf{x}_a^2-\mathbf{x}_b^2}\left(\mathbf{x}_a\,\mathcal{P}^c+\mathbf{x}_b\,\mathcal{S}_c\right)\right)~,
\end{align}
where we have put the gauge-invariant fermions to the left so as to remind the reader that the bosonic derivatives are not meant to act on them in the supercharges. The Hamiltonian $H$ (not $H_m$) in the new parametrization is:
\begin{multline}\label{eq:hamfinal}
H=-\frac{1}{2\Delta}\partial_{\mathbf{x}_a}\Delta\partial_{\mathbf{x}_a}+\frac{1}{2}\left[\mathbf{y}_a\left(\mathcal{P}^{a2}+\mathcal{S}_a^{2}\right)+2\,\mathbf{z}_a\, \mathcal{P}^a\mathcal{S}_a\right]\\+\frac{1}{2}m^2\,\mathbf{x}_a\,\mathbf{x}_a+3g\,m\,\mathbf{x}_1\mathbf{x}_2\mathbf{x}_3+\frac{g^2}{2}\left(\mathbf{x}_1^2\mathbf{x}_2^2+\mathbf{x}_1^2\mathbf{x}_3^2+\mathbf{x}_2^2\mathbf{x}_3^2\right)-\frac{3}{4}m[\bar{\chi}_A,\chi_{A}]+ig\,\epsilon_{AkC}\bar{\chi}_A\,\mathbf{x}_k\,\boldsymbol{\tilde{\sigma}}^k\chi_C~.
\end{multline}

\section{Numerical results}\label{sec:numerics}
\begin{table}[ht!]
\begin{centering}
\begin{tabular}{c c c}
 & ~~~~~$j=0$ &~~~~~$j=1/2$ \\
$R=0$ &\raisebox{-0.55\totalheight}{\includegraphics[width=0.42\textwidth]{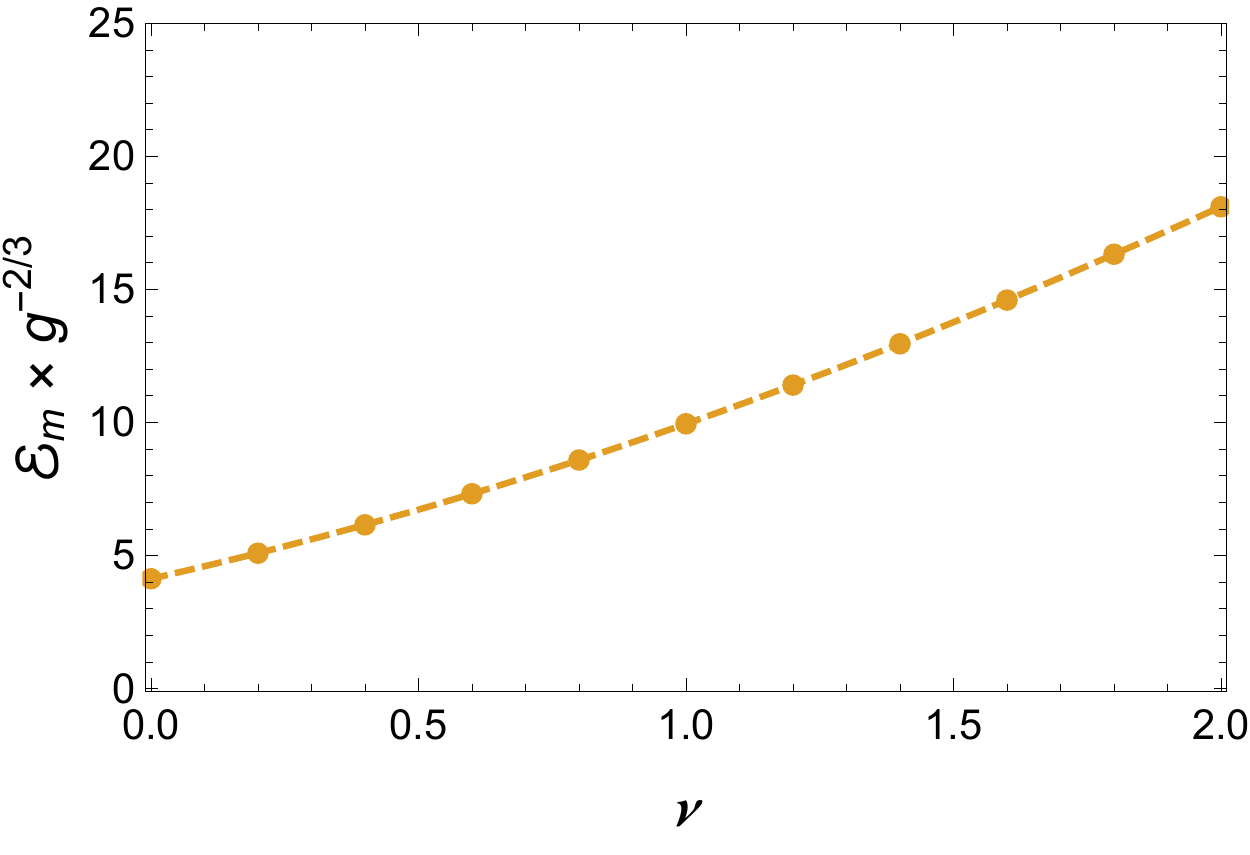}} &\raisebox{-0.55\totalheight}{ \includegraphics[width=0.42\textwidth]{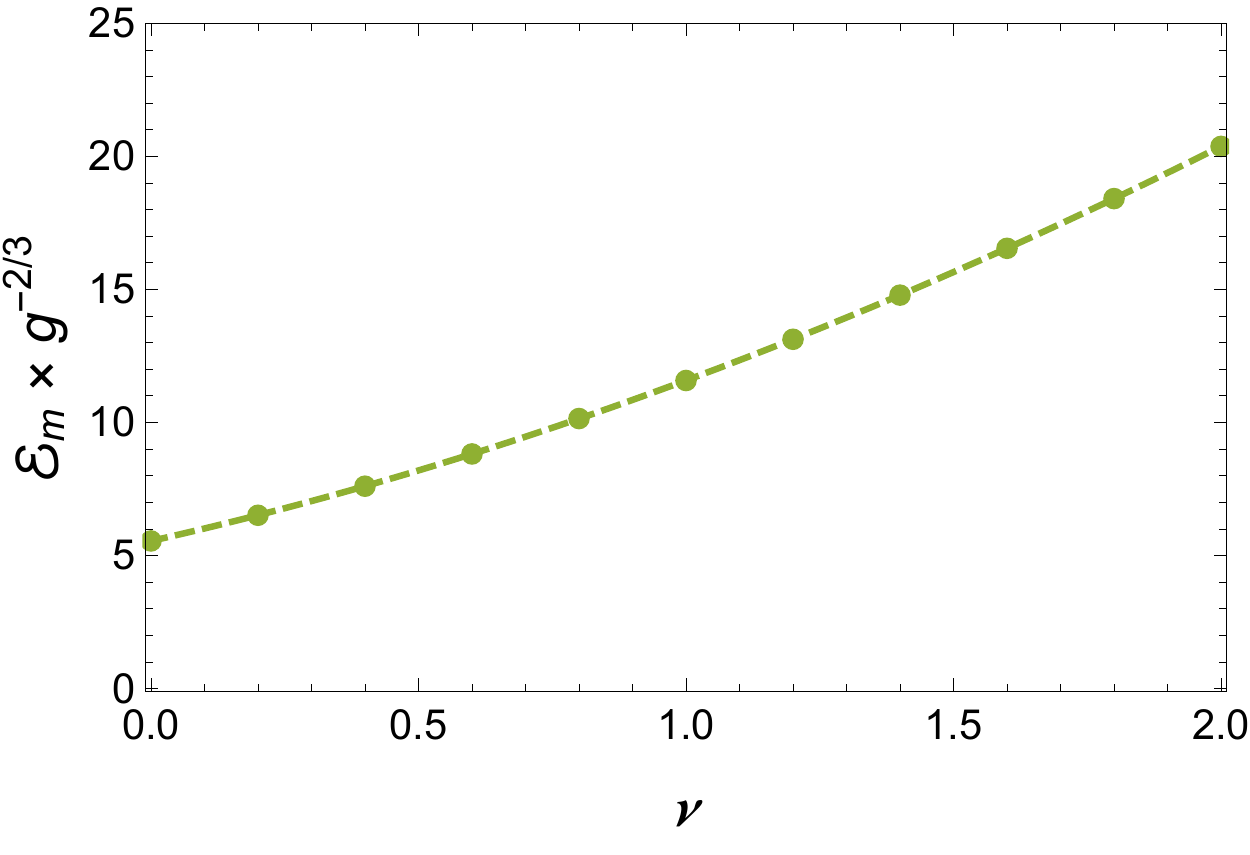} }\\
 $R=1$ &\raisebox{-0.55\totalheight}{\includegraphics[width=0.42\textwidth]{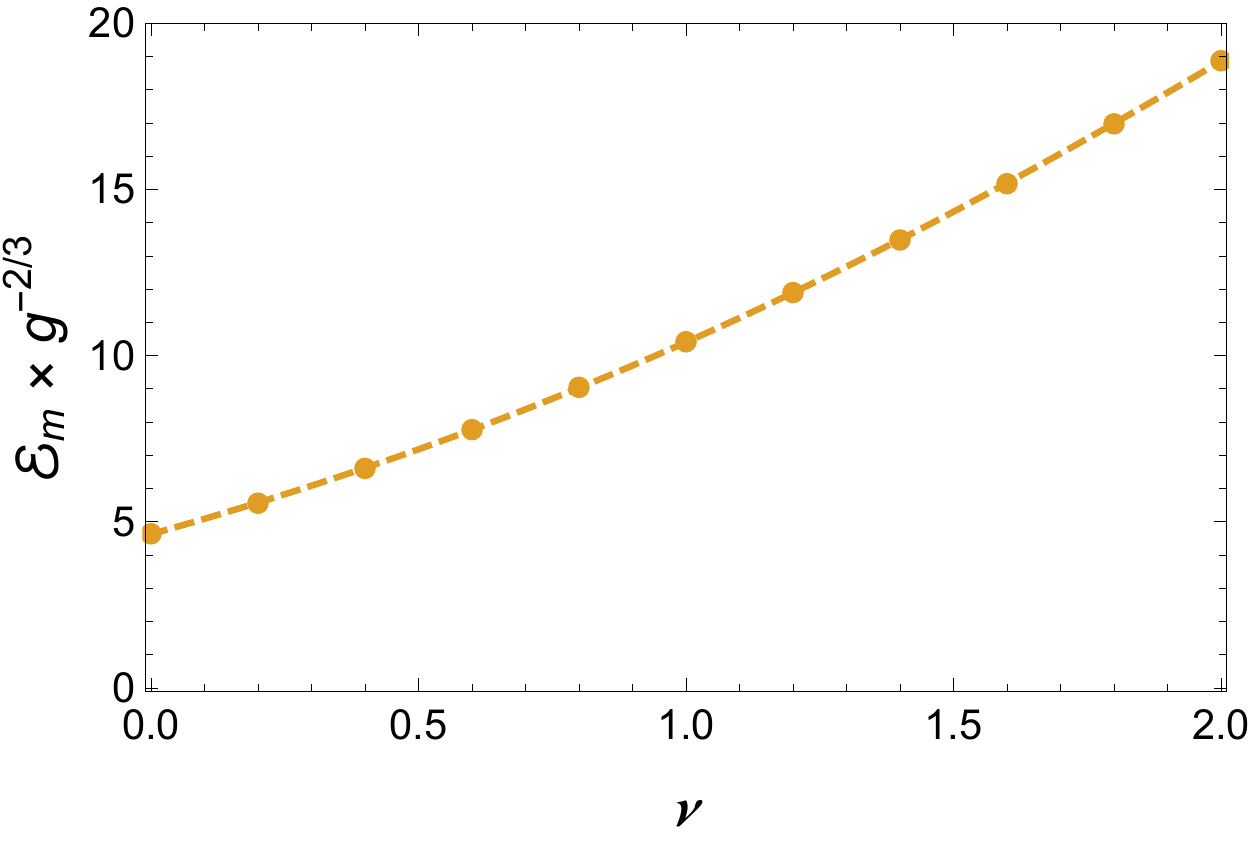}} & \raisebox{-0.55\totalheight}{\includegraphics[width=0.42\textwidth]{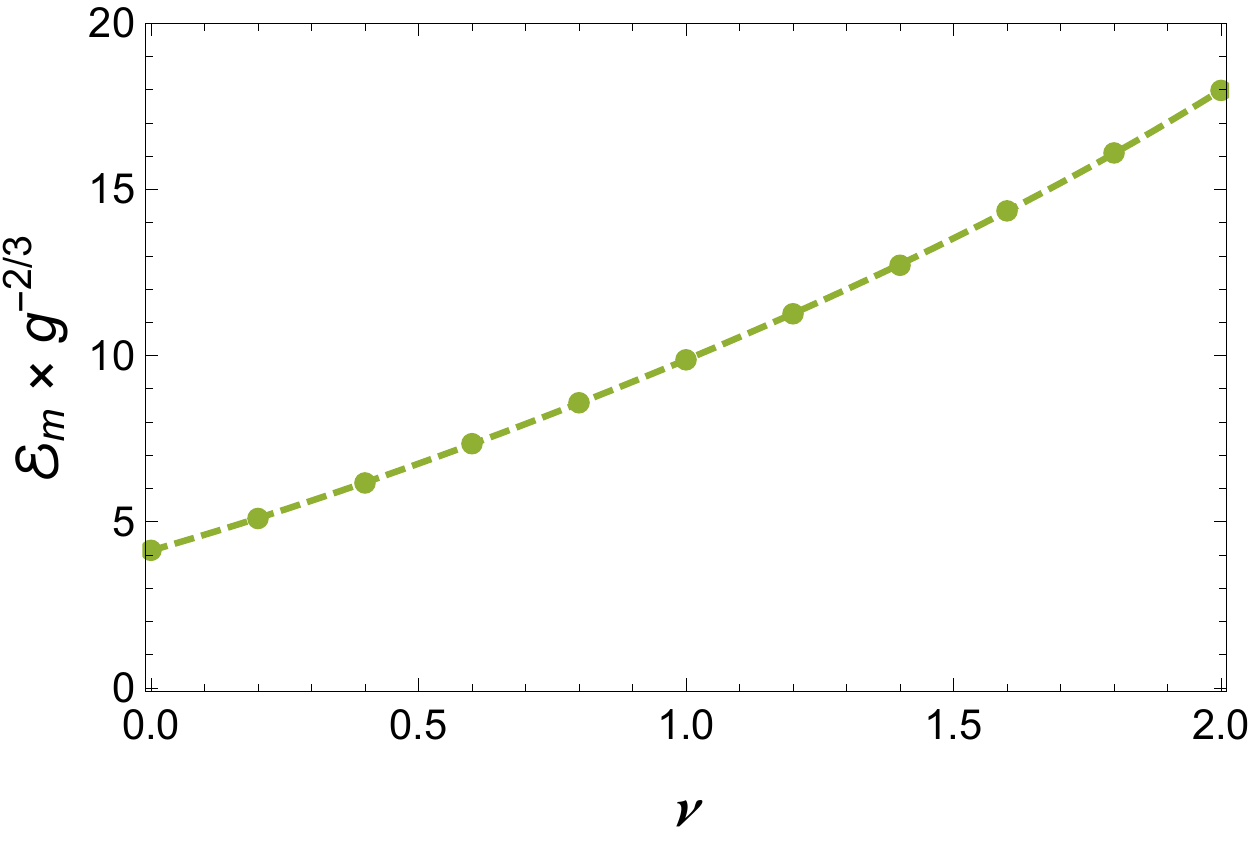}}\\
 $R=2$ &\raisebox{-0.55\totalheight} {\includegraphics[width=0.42\textwidth]{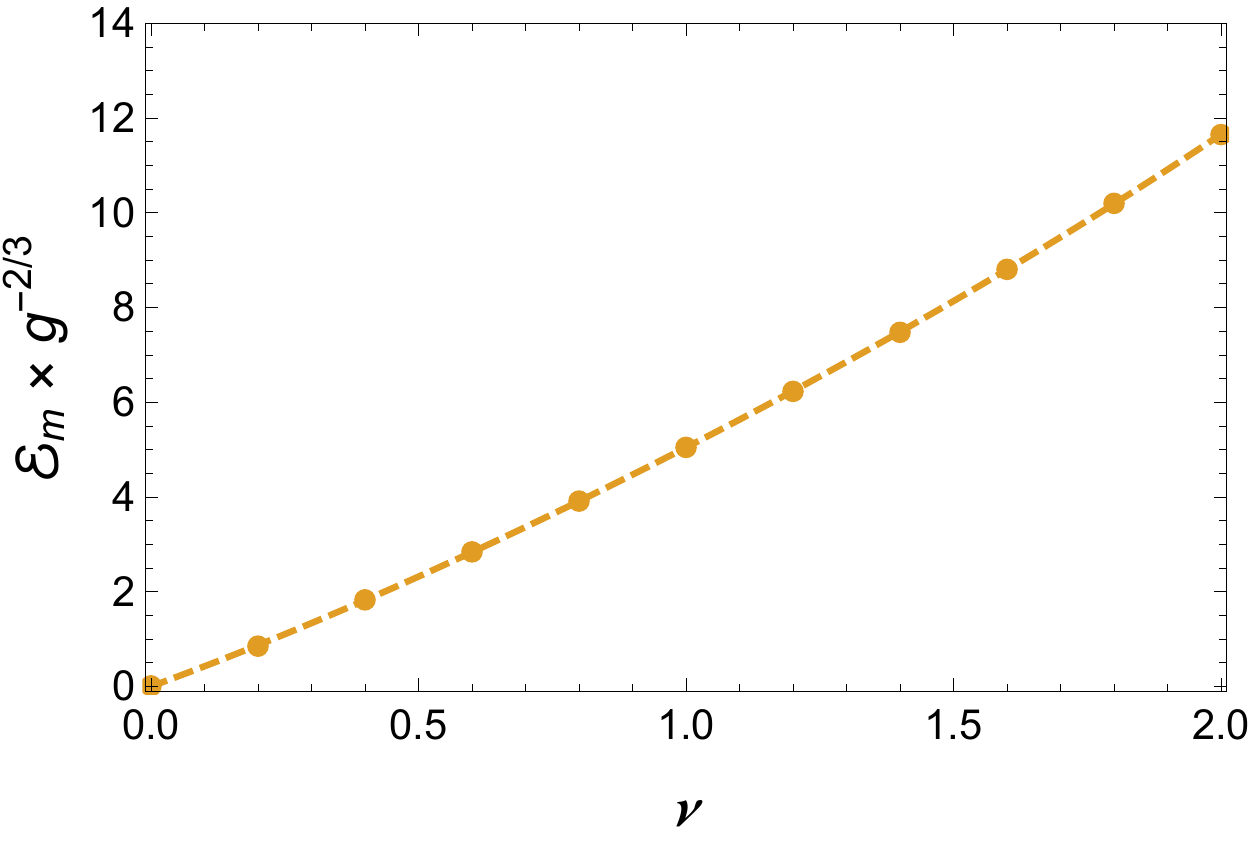}}& \raisebox{-0.55\totalheight}{\includegraphics[width=0.42\textwidth]{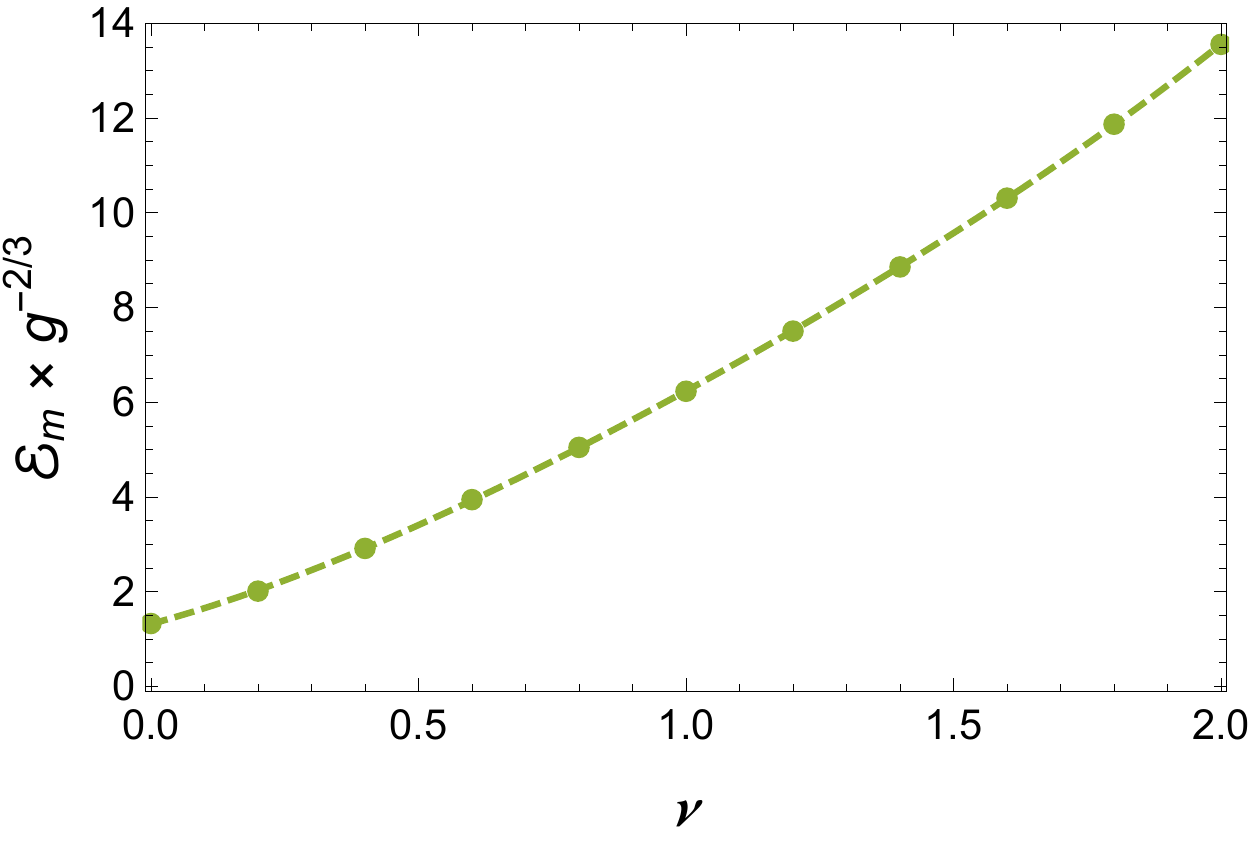}}\\
 $R=3$ &\raisebox{-0.55\totalheight}{\includegraphics[width=0.42\textwidth]{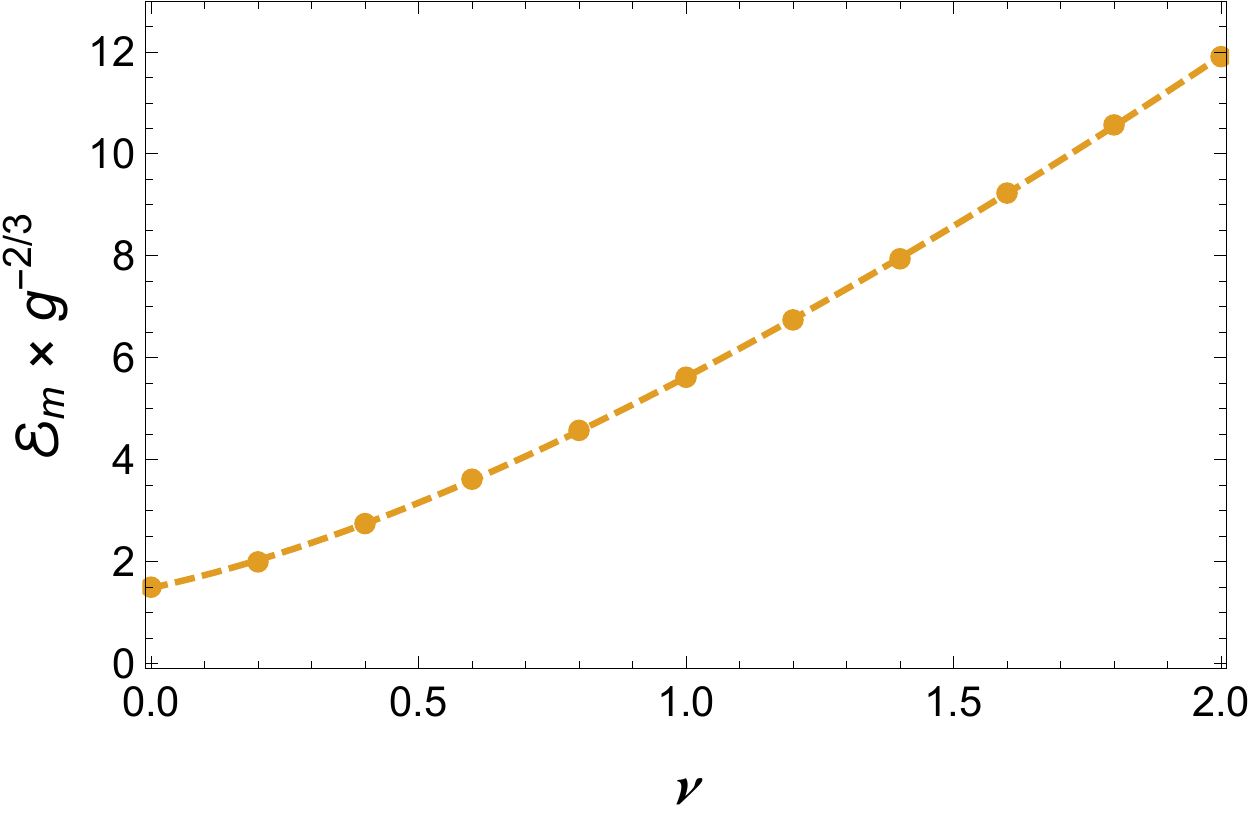}} &\raisebox{-0.55\totalheight}{\includegraphics[width=0.42\textwidth]{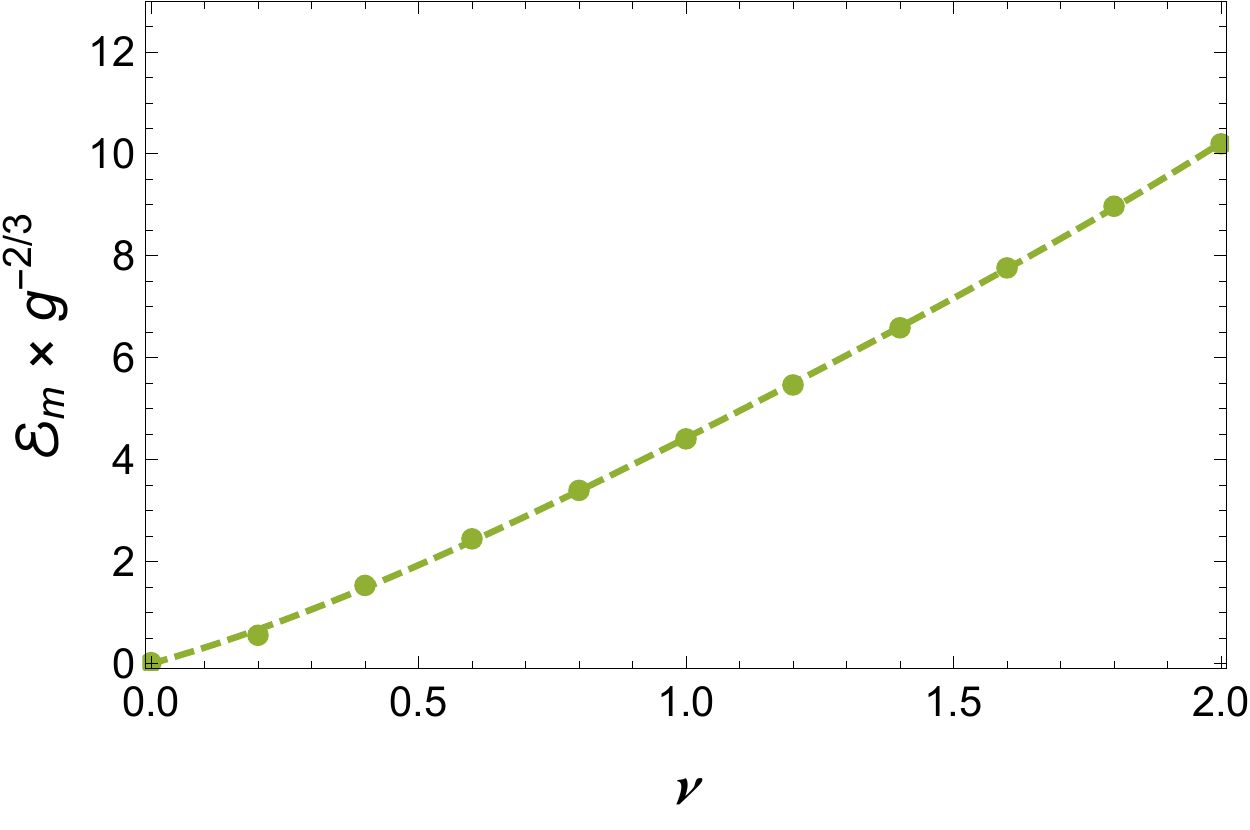}}
\end{tabular}
\end{centering}
\caption{Lowest energy eigenvalue for $R=\{0,1,2,3\}$ and $j=\{0,1/2\}$ as a function of $\nu$. Each row corresponds to a different value of $R$ up to 3 and the columns are labeled by $j=0$ or $j=1/2$. Note that for $\nu=0$ there are $\mathcal{E}=0$ energy eigenstates in both the $R=2$ and $R=3$ sectors of the theory. This implies the existence of 4 supersymmetric ground states at $\nu=0$, a fermionic $j=1/2$ doublet in the $R=3$ sector and two bosonic $j=0$ singlets in the $R=2$ and $R=4$ sectors.}
  \label{tab:results}
\end{table}
In order to calculate the spectrum of the Hamiltonian (\ref{eq:hamfinal}), we must reduce our problem using symmetry, that is we should  label our states via the maximal commuting set of conserved quantities: $H_m,\,J^3\,,\vec{J}^{\,2},\,R$. Because of the discrete particle-hole symmetry (\ref{discretesym}) we need only consider $R=0,\dots,3$. In appendix \ref{ap:schroeq} we construct gauge-invariant highest-weight representations of $SO(3)_J$ in each $R$-charge sector. This means we fix the wavefunctions' dependence on the angles $\vartheta_i$ and $\varphi_i$ and provide the reduced Schr\"{o}dinger operators that depend only on $\mathbf{x}_a$.\footnote{We only provide a small set of these reduced Schr\"{o}dinger operators, as they increase in size with increasing $SO(3)_J$ eigenvalue $j$.}

Our numerical results for the lowest energy states of $H_m$ for each $R$ and $j$ are presented in Table \ref{tab:results} and were obtained by inputting the restricted Schr\"{o}dinger equations of appendix \ref{ap:schroeq} into {\it Mathematica}'s {\tt NDEigenvalues} command, which uses a finite element approach to solve for the eigenfunctions of a coupled differential operator on a restricted domain. We have labeled each row by the fermion number $R$ and each column by the $SO(3)_J$ highest weight eigenvalue $j$ (i.e. $\vec{J}^2|\psi\rangle=j(j+1)|\psi\rangle$ and ${J}^3|\psi\rangle=j|\psi\rangle$).

\begin{figure}
   \begin{centering}
    \includegraphics[width=0.44\textwidth]{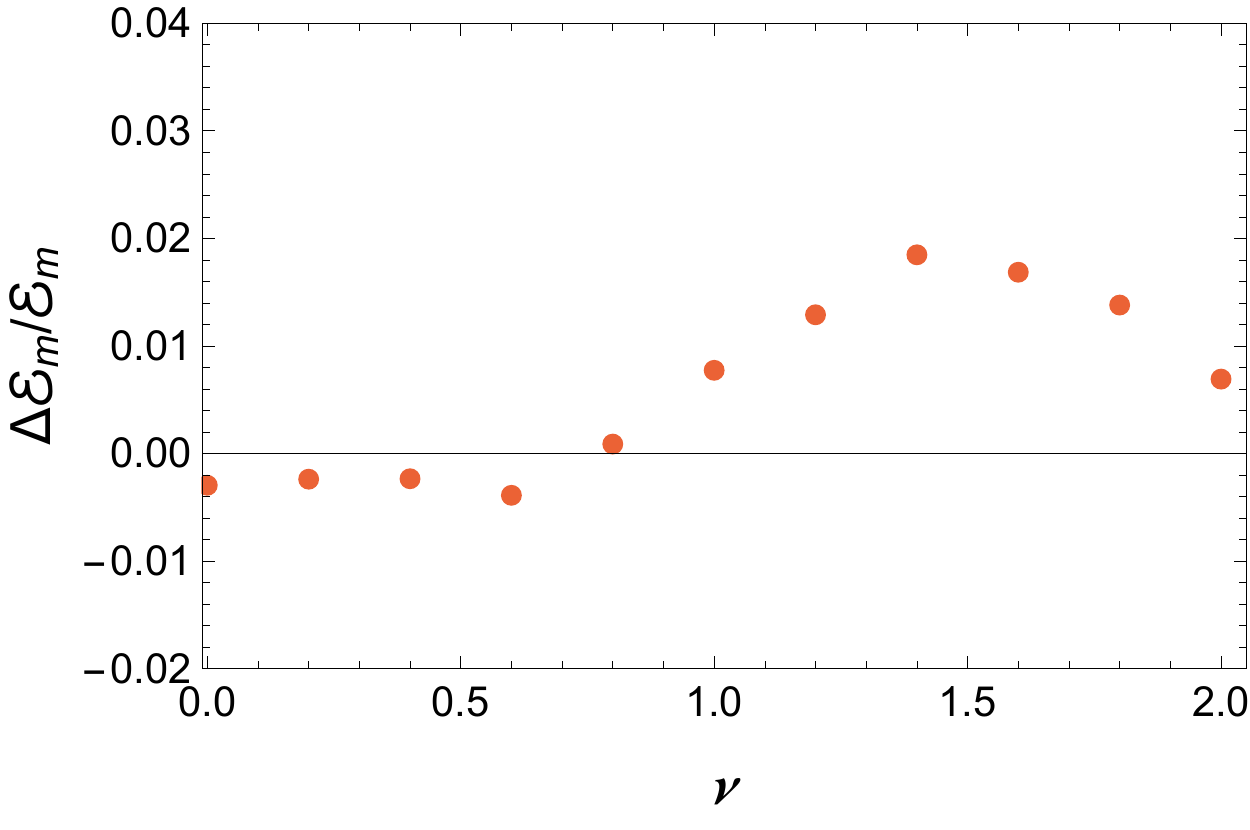}\hfill
   \includegraphics[width=0.44\textwidth]{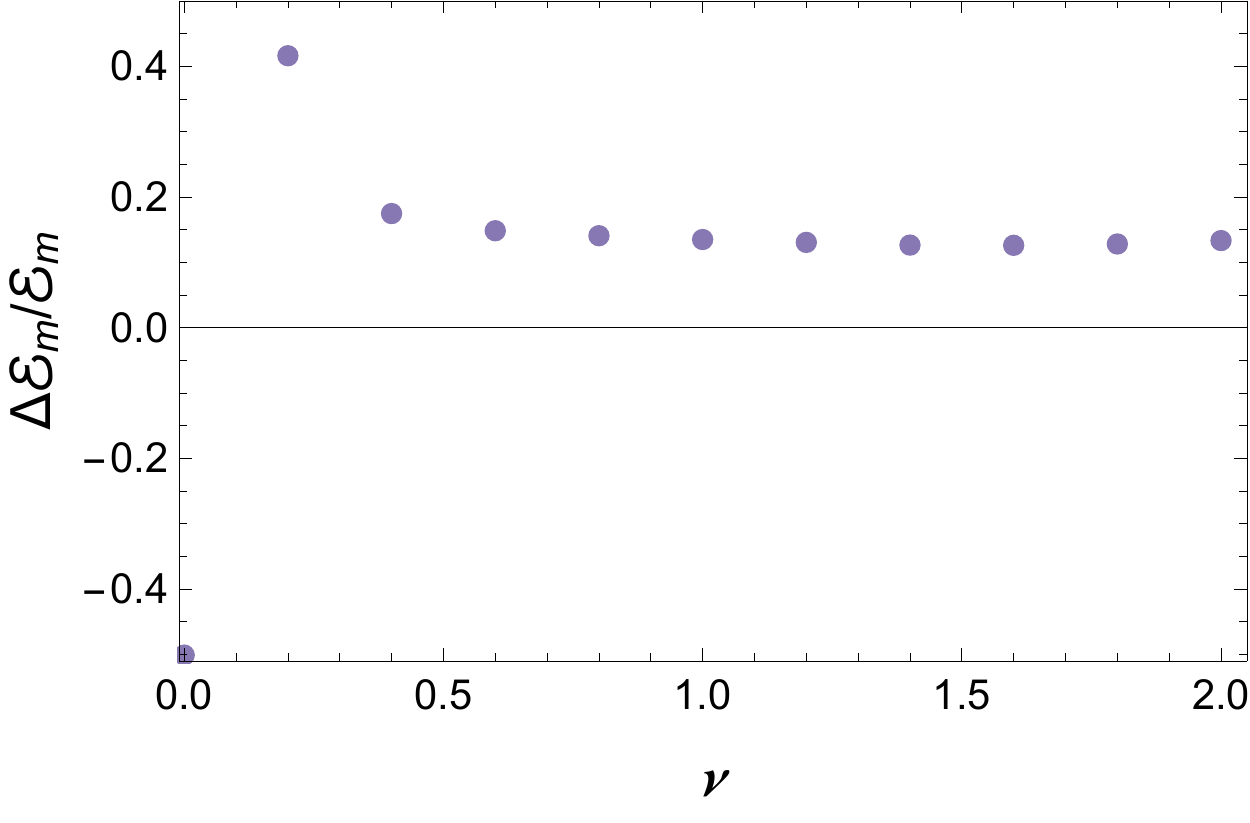}
   \caption{\emph{Left:} Percentage energy difference between the lowest $(R,j)=(0,0)$ state and the lowest $(R,j)=(1,1/2)$ state. As expected by supersymmetry their energies match to within $2\%$. \emph{Right:} Percentage energy difference between the lowest $(R,j)=(2,0)$ state and the lowest $(R,j)=(3,1/2)$ state. These energies match to within $13\%$. The percentage error is larger for low $\nu$ since the energy is approaching zero.}\label{fig:diffplot}
 \end{centering}
 \end{figure}

A few comments are in order:
\begin{enumerate}
\item The most striking feature of these plots is the seeming appearance of zero energy states for $(R,j)=(2,0)$ and $(R,j)=(3,1/2)$ as $\nu\rightarrow0$. Since the Witten index $W_I=0$, and since the states in the $(2,0)$ and $(3,1/2)$ sectors seem have nonzero energy for any finite $\nu$, it must be the case that these states are elements of the same supersymmetry multiplet. This must be so for the deformation invariance of $W_I$.

\item Since we know, by construction, that the lowest energy $(R,j)=(2,0)$ and $(R,j)=(3,1/2)$ states are related by supersymmetry, we can use the difference in their numerically-obtained energies as a benchmark of our numerical errors. Obtaining the $(R,j)=(2,0)$ ground state energy required solving a coupled Schr\"{o}dinger equation involving 15 functions in 3 variables. For the $(R,j)=(3,1/2)$ state, the number of functions one is numerically solving for jumps to 40. In the latter case, it was difficult to reduce our error (either by refining the finite element mesh, or increasing the size of the domain) in a significant way without \emph{Mathematica} crashing. This is despite the fact that we had 12 cores and 64 Gb of RAM at our disposal. In figure \ref{fig:diffplot} we plot the percentage error in the $H_m$ energy difference between these two states as a function of $\nu$. We find that the energy difference between these states is around $13\%$ of the total energy as a function of $\nu$. For comparison, we also do this for the lowest $(R,j)=(0,0)$ and $(R,j)=(1,1/2)$ states, where the numerics are more reliable as a result of solving a much simpler set of equations. There the difference between the computed energies is at most $2\%$.

\item Our results suggest that there are 4 supersymmetric states, two of which are bosonic and two which are fermionic, which would cancel in the evaluation of the index. Explicitly, the two bosonic states are $j=0$ singlets in the $R=2$ and $R=4$ sectors (recall the discrete particle hole symmetry of the theory) and the two fermionic states are the $j=1/2$ doublet in the $R=3$ sector. It is interesting to note that there aren't more states in this multiplet, for example numerically studying the $(R,j)=(0,1)$ sector reveals no evidence for a supersymmetric state in the $\nu\rightarrow 0$ limit.   

\item The massless $SU(2)$ model was studied using a different numerical approach in \cite{Wosiek:2002nm,Campostrini:2004bs} and their plots for the ground state energies seem to approach ours, particularly figures 2 and 5 of \cite{Campostrini:2004bs}.

\item Our numerical evidence for these supersymmetric states does not constitute a proof since we will never be able to numerically resolve if this state has \emph{exactly} zero energy. However, the result is highly suggestive of a supersymmetry preserving set of states at $\nu=0$ and there is no contradiction with the analytically obtained Witten index result $W_I=0$. It would be interesting to analyze the existence of these states analytically in future work.
\end{enumerate}

\section{Effective theory on the moduli space of the $SU(2)$ model}\label{sec:effectiv}
In order to get a better handle on the previous section's numerical results, we will now study the $\nu\rightarrow0$ limit of the matrix model analytically. Since the full problem is clearly quite difficult even for $N=2$, we will study the massless model in some parametric limit. This is possible because the theory has a \emph{moduli space}\footnote{Also sometimes called a \emph{Coulomb branch}.}---a flat direction where the D-branes can become well separated, and along this moduli space certain fields become massive and can be integrated out. We will parametrize this moduli space by the coordinates $(\mathbf{x}_3,\vartheta_2,\vartheta_1)$ and will henceforth label them $(\mathbf{x}_3,\vartheta_2,\vartheta_1)\rightarrow(r,\theta,\phi)$ for the remainder of this section. The parametric limit we will take is the limit of large $r$.

To derive the effective theory along the moduli space we will first take $(r,\theta,\phi)$ to be slowly varying and expand $H=H^{(0)}+H^{(1)}+\dots$ in inverse powers of the dimensionless quantity $g\,r^3$. We will compute the effective Hamiltonian in perturbation theory by integrating out the other fields in their ground state, in which $(r,\theta,\phi)$ appear as parameters. Similar analysis to this was performed in \cite{Halpern:1997fv,Smilga:1989ew,Smilga:2002mra,Akhmedov:2002mi,Frohlich:1999zf,Graf:1998bm}.  Defining $\vec{\partial}\equiv\left(\partial_{\mathbf{x}_1},\partial_{\mathbf{x}_2}\right)$, the Hamiltonian, to lowest order, is
\begin{multline}
H^{(0)}\equiv-\frac{1}{2\left(\mathbf{x}_1^2-\mathbf{x}_2^2\right)}\vec{\partial}\cdot\left(\mathbf{x}_1^2-\mathbf{x}_2^2\right)\vec{\partial}-\frac{1}{2\left(\mathbf{x}_1^2-\mathbf{x}_2^2\right)^2}\left[\left(\mathbf{x}_1^2+\mathbf{x}_2^2\right)\left(\partial_{\vartheta_3}^2+\partial_{\varphi_3}^{2}\right)+4\,\mathbf{x}_1\,\mathbf{x}_2\, \partial_{\vartheta_3}\partial_{\varphi_3}\right]\\+\frac{g^2}{2}r^2\left(\mathbf{x}_1^2+\mathbf{x}_2^2\right)-i\,g\,r\,\epsilon_{3DE}\,\bar{\chi}_D\,\tilde{\boldsymbol{\sigma}}^3\,\chi_E~,
\end{multline}
where $\tilde{\boldsymbol{\sigma}}^{i~\beta}_\alpha\equiv M^{ji}\boldsymbol{\sigma}^{j~\beta}_\alpha$ depends explicitly on $(\vartheta_1,\vartheta_2,\vartheta_3)$. It is straightforward to show that $H^{(0)}$ admits a zero energy ground state given by:
\begin{equation}
  \Psi^{(0)}=\frac{g\,r}{\pi\sqrt{32}} e^{-\frac{g}{2}\,r\,\left(\mathbf{x}_1^2+\mathbf{x}_2^2\right)}\sum_{B=1}^2\left\lbrace\bar{\chi}_B\,\epsilon\,\bar{\chi}_B-i\sum_{C=1}^2\epsilon_{3BC}\,\bar{\chi}_B\,\tilde{\boldsymbol{\sigma}}^3\left(\bar{\chi}_C\epsilon\right)\right\rbrace|0\rangle~,
\end{equation}
where $|0\rangle$ is the fermionic vacuum and we have normalized $\Psi^{(0)}$ with respect to
\begin{equation}
  \int_0^\infty d\mathbf{x}_1\int_{-\mathbf{x}_1}^{\mathbf{x}_1}d\mathbf{x}_2\int_0^{2\pi}d\vartheta_3\int_0^{2\pi}d\varphi_3\,\left(\mathbf{x}_1^2-\mathbf{x}_2^2\right)~.
\end{equation}
Similarly we can expand the supercharges $Q_\alpha= Q_\alpha^{(0)}+Q_\alpha^{(1)}+\dots$, where
\begin{align}
Q^{(0)}_\alpha&\equiv -i\sum_{a,b=1}^2\tilde{\boldsymbol{\sigma}}_\alpha^{b~\gamma}\chi_{a\gamma}\left(\delta_{ab}\left\lbrace\partial_{\mathbf{x}_b}+\frac{g}{2}|\epsilon_{bst}|\,\mathbf{x}_s\mathbf{x}_t\right\rbrace+i\frac{\epsilon_{abc}}{\mathbf{x}_a^2-\mathbf{x}_b^2}\left(\mathbf{x}_a\,\mathcal{P}^c+\mathbf{x}_b\,\mathcal{S}_c\right)\right)~,\\
Q^{(1)}_\alpha&\equiv-i\,\tilde{\boldsymbol{\sigma}}_\alpha^{b~\gamma}\chi_{3\gamma}\left(\delta_{3b}\left\lbrace\partial_r+\frac{g}{2}|\epsilon_{3st}|\,\mathbf{x}_s\mathbf{x}_t\right\rbrace+i\frac{\epsilon_{3bc}}{r}\left(\mathcal{P}^c+\mathcal{S}_c\right)\right)~.
\end{align}
It is easy to check that $Q^{(0)}_\alpha\Psi^{(0)}=\bar{Q}^{(0)\beta}\Psi^{(0)}=0$. We are now tasked with finding the effective supercharges $Q^{\rm eff}_\alpha= \left\langle Q^{(1)}_\alpha\right\rangle_{\Psi^{(0)}}+\dots$ that act on the massless degrees of freedom $(r,\theta,\phi)$ along the moduli space. At lowest order we find the supercharges (acting on gauge-invariant wavefunctions) are those of a free particle in $\mathds{R}^3$ and its fermionic superpartner:
\begin{equation}
Q_\alpha^{\rm eff}=-i\,\nabla_{\vec{\mathbf{x}}}\cdot\vec{\boldsymbol{\sigma}}_\alpha^{~\,\gamma}\,\psi_\gamma~,\quad\quad\quad
\bar{Q}^{\beta}_{\rm eff}=-i\,\bar{\psi}^\gamma\nabla_{\vec{\mathbf{x}}}\cdot\vec{\boldsymbol{\sigma}}_\gamma^{~\,\beta}~,
\end{equation}
where we have labeled $(r,\theta,\phi)$ in cartesian coordinates as well as defined $\left(\psi_\alpha,\bar{\psi}^\beta\right)\equiv\left(\chi_{3 \alpha},\bar{\chi}_3^\beta\right)$. Since the remaining gauge angles $(\varphi_1,\varphi_2)$ have no kinetic terms in the effective theory along the moduli space, we need not consider them as dynamical variables and can treat $\psi_\alpha$ as a fundamental field.

Let us now compute the effective theory to next order in perturbation theory. Instead of computing this in the operator formalism, let us first invoke symmetry arguments to constrain what the answer should look like. The low energy effective theory on the moduli space should be a supersymmetric theory with four supercharges and an $SO(3)$ $R-$symmetry, therefore it should fall in the class discovered in \cite{Ivanov:1990jn,Diaconescu:1997ut}:
\begin{multline}\label{eq:efac}
\mathcal{L}=\frac{1}{2}f\left(\dot{\vec{\mathbf{x}}}^2+i(\bar{\psi}\dot{\psi}-\dot{\bar{\psi}}\psi) +D^2\right)+\frac{1}{2}\left(\nabla_k f\right)\,\epsilon_{klm}\,\dot{\mathbf{x}}^l\,\bar{\psi}\,\boldsymbol{\sigma}^m\,\psi\\-\frac{D}{2}\,\left(\nabla_{\vec{\mathbf{x}}}f\right)\cdot\bar{\psi}\,\vec{\boldsymbol{\sigma}}\,\psi+\frac{1}{4}\left(\nabla_i\nabla_j f\right)\left(\bar{\psi}\,\boldsymbol{\sigma}^i\,\psi\right)\left(\bar{\psi}\,\boldsymbol{\sigma}^j\,\psi\right)~,
\end{multline}
which is invariant under
\begin{align}
\delta\vec{\mathbf{x}}&=i\bar{\psi}\,\vec{\boldsymbol{\sigma}}\,\xi-i\bar{\xi}\,\vec{\boldsymbol{\sigma}}\,\psi\nonumber\\
\delta\psi_\alpha&=\dot{\vec{\mathbf{x}}}\cdot\vec{\boldsymbol{\sigma}}_\alpha^{~\,\beta}\xi_\beta+iD\,\xi_\alpha\nonumber\\
\delta\bar{\psi}^\beta&=\dot{\vec{\mathbf{x}}}\cdot\bar{\xi}^\alpha\vec{\boldsymbol{\sigma}}_\alpha^{~\,\beta}-iD\,\bar{\xi}^\beta\nonumber\\
\delta D&=-\dot{\bar{\psi}}\,\xi-\bar{\xi}\,\dot{\psi}~.\nonumber
\end{align}
In order to preserve the $SO(3)$ symmetry $f$ should be a function of $r\equiv |\vec{\mathbf{x}}|$. Notice that (\ref{eq:efac}) reduces to the theory of a free particle and its superpartner when $f=1$. Therefore we should find that at 1-loop order $f=1+\frac{c}{g\,r^3}$, since $\left(g\,r^3\right)^{-1}$ is our expansion parameter, with $c$ to be determined. A calculation \cite{Smilga:2002mra,Akhmedov:2002mi} reproduced in appendix \ref{ap:metric} gives $c=-3/2$ or
\begin{equation}\label{eq:modulimet}
  f=1-\frac{3}{2g\,r^3}~.
\end{equation}
Analytic evidence for the numerically found supersymmetric ground states can be obtained by studying the Schr\"{o}dinger problem associated with (\ref{eq:efac}). We do not do this here, but we can gain some intuition by studying the existence of normalizable zero-modes of the Laplacian on moduli-space \cite{Douglas:1996yp}:
\begin{equation}
  ds^2=\left(1-\frac{3}{2g\,r^3}\right)\left(dr^2+r^2\,d\Omega_2^2\right)~.
\end{equation}
We can construct two normalizable zero-modes as follows. The zero-form
\begin{equation}
  \omega_0\equiv\int^r dr'\,\frac{1}{r'^{2}}\left(1-\frac{3}{2g\,r'^3}\right)^{-1/2}
\end{equation}
is a zero-mode of the Laplacian, but is not normalizable. To construct normalizable forms, we take
\begin{equation}
  \omega_1\equiv d\omega_0~,\quad\quad \omega_2=\star\,\omega_1~.
\end{equation}
These are normalizable within the domain $r\in \left[\left(\frac{3}{2g}\right)^{1/3},\infty\right]$. Since there exists zero-modes in this toy-moduli-space approximation, it would be interesting to study the set of ground states of (\ref{eq:efac}) in more detail.

\section{Discussion}\label{sec:discussion}
In this paper we have studied the mini-BFSS/BMN model with gauge group $SU(2)$ and uncovered numerical evidence for a set of supersymmetric ground states in the massless limit of the theory. In the massless limit the matrices can become widely separated. The effective theory on the moduli space has non-trivial interactions governed by a metric that gets generated on this moduli space at one loop. Let us now discuss what may happen in the $SU(N)$ case at large $N$. The quartic interaction in (\ref{eq:fullhamiltonian}) can be rewritten as a commutator-squared interaction $\left(f_{ABC}\,X^i_B\,X^j_C\right)^2\sim \text{Tr}\left(\left[X^i,X^j\right]\right)^2$, where $X^i\equiv X^i_A\,\tau_A$ and $\tau_A$ are the generators in the fundamental of $SU(N)$. Therefore, at tree-level, along the moduli space there will be a set of $N-1$ massless, non-interacting, point particles in $\mathds{R}^3$ (and their superpartners), each one corresponding to an element of the Cartan of $SU(N)$. At one-loop there will be a correction to the moduli space-metric, depending on the relative distances between these particles. Just like in the $SU(2)$ case these corrections will come at order $|r_a-r_b|^{-3}$. One difference, however, is that there may be an enhancement of order $N$ to this correction. It would certainly be interesting to see if we can isolate the $|r_a-r_b|^{-3}$ corrections to the moduli space metric by taking a large $N$ limit, as can be done in the D0-D4 system \cite{Douglas:1996yp} and in the three-node Abelian quiver \cite{Anninos:2013nra}. Perhaps we can adapt the methods in \cite{Lin:2014wka} for these purposes. The analysis in \cite{Smilga:2002mra} seems to suggest that such a decoupling limit at large $N$ is possible .

Interestingly, it was shown in \cite{Anninos:2013nra} that the one-loop effective action on the Coulomb branch of the three-node Abelian quiver exhibits an emergent conformal symmetry at large $N$. This conformal symmetry depends on the delicate balance between the form of the interaction potential and the metric on the moduli space, which has a similar $|r_a-r_b|^{-3}$ form as in (\ref{eq:modulimet}). It would be interesting to establishing whether the $SU(N)$ generalization of the model studied in this paper also has a non-trivial conformal symmetry at infinite $N$, broken by finite $N$ effects. We save this problem for future work, but list here some reasons why this would be worth studying:
\begin{enumerate}
\item The BFSS matrix model has a holographic interpretation \cite{Taylor:2001vb,Susskind:1998vk,Balasubramanian:1997kd,Polchinski:1999br}. At large $N$ it is dual to a background of D0 branes in type IIA supergravity. In BFSS there is no correction to the moduli space metric and neither side of this duality is conformal. The BFSS matrix model is thus a theory of the 10d flat space S-matrix. It would be interesting to understand the large $N$ version of mini-BFSS in the context of holography along similar lines. Because of the large number of coupled degrees of freedom at large $N$ and the reduced supersymmetry, the effective theory along the moduli space of mini-BFSS has a non-trivial metric and may potentially exhibit a non-trivial conformal fixed point along this moduli space, as happens for quiver quantum mechanics models with vector rather than matrix interactions \cite{Anninos:2013nra,Anninos:2016szt}. To answer this question definitively we will need to compute the effective theory along the Coulomb branch for $N\gg 2$ and check whether it is conformal. 

\item New results have shown that a certain class of disordered quantum mechanics models, known as SYK for Sachdev-Ye-Kitaev, exhibits phenomenology of interest for near-extremal black holes (see \cite{KitaevTalks,Anninos:2016szt,Polchinski:2016xgd,Maldacena:2016hyu,Jensen:2016pah,Sachdev:2015efa,Jevicki:2016bwu,Cotler:2016fpe} and references therein as well was \cite{Witten:2016iux,Gurau:2016lzk,Klebanov:2016xxf,Krishnan:2016bvg} for models without disorder).  These phenomena include an emergent conformal symmetry in the IR, maximal chaos \cite{Maldacena:2015waa}, and a linear in $T$ specific heat. Despite the successes of these models, they are not dual to weakly coupled gravity. BFSS is a large $N$ gauged matrix quantum mechanics dual to weakly-coupled Einstein gravity but, as we previously mentioned, it does not have an emergent conformal symmetry and remains a model of D-particles in flat space. It would certainly be interesting if mini-BFSS fell in the universality class of quantum mechanics models with emergent conformal symmetry in the IR and maximal chaos, such as the SYK model and its non-disordered cousins, but remains dual to weakly coupled gravity. Recently \cite{Anninos:2016klf,Ferrari:2017ryl} advocated the study of such matrix models for similar reasons. In the same vein \cite{Gur-Ari:2015rcq} studies classical chaos in BFSS numerically.

\item If this model, like SYK, is at all related to the holography of near-extremal black holes, then we can try to study its $S$-matrix to gain some insight into the real time dynamics of black hole microstates. A numerical implementation of such a study in the context of similar supersymmetric quantum mechanics models with flat directions can be found in \cite{Balthazar:2016utu}.

\item The slow moving dynamics of a class of BPS multi-black hole solutions in supergravity is a superconformal quantum mechanics \cite{Ferrell:1987gf,Michelson:1999zf,Michelson:1999dx,BrittoPacumio:1999ax} with no potential, provided a near horizon limit is taken. It would be interesting to understand if there is some limit in which the multi-black hole moduli space quantum mechanics and the large $N$ matrix quantum mechanics on the moduli space coincide. Perhaps as a consequence of non-renormalization theorems as in \cite{Denef:2002ru}.

\end{enumerate}


\section*{Acknowledgements}
It is a pleasure to thank Dionysios Anninos, Frederik Denef, Felix Haehl, Rachel Hathaway, Eliot Hijano, Jaehoon Lee, Eric Mintun, Edgar Shaghoulian, Benson Way and Mark Van Raamsdonk for helpful discussions. We are particularly indebted to Dionysios Anninos, Frederik Denef, and Edgar Shaghoulian for their comments on an early draft. We made heavy use of Matthew Headrick's {\tt grassmann.m} package. C.C. would like to thank David and Gay Cogburn for their support. T.A. is supported in part by the U.S. Department of Energy under grant Contract Number DE-SC0012567, by the Natural Sciences and Engineering Research Council of Canada, and by grant 376206 from the Simons Foundation.

\appendix

\section{Reduced Schr\"{o}dinger equation}\label{ap:schroeq}
In this appendix we construct gauge-invariant highest-weight wavefunctions of $SO(3)_J$ in each $R-$charge sector (up to 3) and use these to maximally reduce the Schr\"{o}dinger equation via symmetries.
\subsection{R=0}
This sector of the theory was studied in \cite{Asatrian:1982ga,Savvidy:1982wx,Savvidy:1984gi}, although without access to numerics. We repeat their analysis here. We wish to separate variables using the $SO(3)_J$ symmetry. We therefore want to write down the highest weight state satisfying $J^3|\psi\rangle_0=j|\psi\rangle_0$ and $J^{+}|\psi\rangle_0=0$~, with $J^{\pm}\equiv J^1\pm i\,J^2$. The rest of the spin multiplet can be obtained by acting on $|\psi\rangle_0$ with $J^{-}$ up to $2j$ times. This however doesn't entirely fix the angular dependence of the wavefunction, as these two conditions only fix the dependence on up to two angles. Recall, however, that  the operators $\vec{P}$ commute with $\vec{J}$ and $\vec{P}^2=\vec{J}^2$, but $\left[H,\vec{P}\right]\neq 0$. We will then write $|\psi\rangle_0$ as a sum of terms with definite $P^3$ eigenvalue. That is, we write $|\psi\rangle_0$ as:
\begin{equation}
|\psi\rangle_0=e^{i\,j\,\vartheta_1}\sin^j\vartheta_2\sum_{p=-j}^j e^{i\,p\,\vartheta_3}\cot^p\left(\frac{\vartheta_2}{2}\right)\,f^p(\mathbf{x}_a)~.
\end{equation}
Since the number of terms in the wavefunction grows with $j$ it will be cumbersome to give the reduced radial Schr\"{o}dinger equation for arbitrary $j$. Instead we will give the expressions for $j=0,\frac{1}{2},1$.
Before giving the reduced Schr\"{o}dinger equations it is worth noting that it has long been known that there exists no supersymmetric states in this sector \cite{Claudson:1984th}. The reason is that the supersymmetry equations $Q_\alpha|\psi\rangle_0=\bar{Q}^\beta|\psi\rangle_0=0$ are easy to solve and give
\begin{equation}
|\psi\rangle_0^{\rm SUSY}\sim\exp\left\lbrace g\, \mathbf{x}_1\,\mathbf{x}_2\,\mathbf{x}_3+\frac{m}{2}\mathbf{x}_a\,\mathbf{x}_a\right\rbrace~,
\end{equation}
which is non-normalizable. It is also known that the spectrum in this sector is discrete \cite{Savvidy:1984gi}.

For parsimony let us define
\begin{equation}
\hat{\mathcal{H}}\equiv-\frac{1}{2\Delta}\partial_{\mathbf{x}_a}\Delta\partial_{\mathbf{x}_a}+V
\end{equation}
with $V$ defined in (\ref{pot}).
Then for $j=0$ the reduced Schr\"{o}dinger equation, obtained from $H_m|\psi\rangle_0=\mathcal{E}_m|\psi\rangle_0$, is simply
\begin{equation}
\left(\hat{\mathcal{H}}+\frac{9}{2}m\right)f^0\left(\mathbf{x}_a\right)=\mathcal{E}_m\,f^0\left(\mathbf{x}_a\right)~.
\label{eq:r0j0SE}
\end{equation}
For $j=1/2$ there is no mixing between the $f^{\pm1/2}\left(\mathbf{x}_a\right)$ and each satisfies
\begin{equation}
\left(\hat{\mathcal{H}}+\frac{9}{2}m+\frac{T}{8}\right)f^{\pm1/2}\left(\mathbf{x}_a\right)=\mathcal{E}_m\,f^{\pm1/2}\left(\mathbf{x}_a\right)~,
\label{eq:r0j12SE}
\end{equation}
where $T$ was defined in (\ref{eq:Tdef}). Finally, for $j=1$ we have
\begin{equation}
\left\lbrace\hat{\mathcal{H}}+\frac{9}{2}m+\frac{T}{4}+\frac{1}{4}\begin{pmatrix}\mathbf{y}_3&0 &\mathbf{y}_1-\mathbf{y}_2\\
0 &T-2\,\mathbf{y}_3&0\\
\mathbf{y}_1-\mathbf{y}_2&0 &\mathbf{y}_3\end{pmatrix}\right\rbrace\begin{pmatrix}f^{-1}\\ f^0\\ f^{+1}\end{pmatrix}=\mathcal{E}_m\begin{pmatrix}f^{-1}\\ f^0\\ f^{+1}\end{pmatrix}~.
\label{eq:r0j1SE}
\end{equation}

\subsection{R=1}
Continuing on from the last section, we want to write down wavefunctions in the $R=1$ sector that are gauge invariant, and satisfy $J^3|\psi\rangle_1=j|\psi\rangle_1$ and $J^+|\psi\rangle_1=0$. To do so, we will write our wavefuntions as a
\begin{equation}
|\psi\rangle_1=e^{i\,j\,\vartheta_1}\sin^j\vartheta_2\sum_{p=-j}^j e^{i\,p\,\vartheta_3}\cot^p\left(\frac{\vartheta_2}{2}\right)\,f^{p}_{A\alpha}\,\bar{\chi}_A^\alpha|0\rangle~,
\end{equation}
where $|0\rangle$ is the fermionic vacuum and each term in the sum has definite $P^3$ eigenvalue.
The functions $f^{p}_{A\alpha}$ that satisfy these conditions are:
\begin{align}
f^{p}_{A1}&=e^{-i\frac{\vartheta_1}{2}}\left\lbrace e^{-i\frac{\vartheta_3}{2}}\cos\left(\frac{\vartheta_2}{2}\right)L^p_{2A-1}\left(\mathbf{x}_a\right)-\,e^{i\frac{\vartheta_3}{2}}\sin\left(\frac{\vartheta_2}{2}\right)L^p_{2A}\left(\mathbf{x}_a\right)\right\rbrace~, \\
 f^{p}_{A2}&=~~e^{i\frac{\vartheta_1}{2}}\left\lbrace e^{-i\frac{\vartheta_3}{2}}\sin\left(\frac{\vartheta_2}{2}\right)L^p_{2A-1}\left(\mathbf{x}_a\right)+\,e^{i\frac{\vartheta_3}{2}}\cos\left(\frac{\vartheta_2}{2}\right)L^p_{2A}\left(\mathbf{x}_a\right)\right\rbrace~.
\end{align}
We remind the reader that the $\bar{\chi}_A^\alpha$ are the gauge-invariant fermions defined in (\ref{gaugeinvferm}). The reduced Schr\"{o}dinger equation for $j=0$ (and hence $p=0$) is
\begin{equation}
\left\lbrace\hat{\mathcal{H}}+\frac{7}{2}m+\frac{5}{8}T+\mathbf{A}
\right\rbrace\begin{pmatrix}L_1^0\\\vdots\\L_6^0\end{pmatrix}=\mathcal{E}_m\begin{pmatrix}L_1^0\\\vdots\\L_6^0\end{pmatrix}
\label{eq:r1j0SE}
\end{equation}
where $\mathbf{A}$ is a $6\times 6$ matrix that can be written in terms of $2\times 2 $ blocks as follows
\begin{equation}
\mathbf{A}\equiv\frac{i}{2}\begin{pmatrix}
i\,\mathbf{y}_1\mathbb{1} &-\left(2 g\,\mathbf{x}_3+\mathbf{z}_3\right)\boldsymbol{\sigma}^3 &~~~\left(2 g\,\mathbf{x}_2+\mathbf{z}_2\right)\boldsymbol{\sigma}^2\\
~~~\left(2 g\,\mathbf{x}_3+\mathbf{z}_3\right)\boldsymbol{\sigma}^3 &i\,\mathbf{y}_2\mathbb{1} &-\left(2 g\,\mathbf{x}_1+\mathbf{z}_1\right)\boldsymbol{\sigma}^1\\
-\left(2 g\,\mathbf{x}_2+\mathbf{z}_2\right)\boldsymbol{\sigma}^2 &~~~\left(2 g\,\mathbf{x}_1+\mathbf{z}_1\right)\boldsymbol{\sigma}^1 &i\,\mathbf{y}_3\mathbb{1}
\end{pmatrix}~,
\end{equation}
where the coordinates $\mathbf{y}_a$ and $\mathbf{z}_a$ (nonlinearly related to $\mathbf{x}_a$) were defined in (\ref{nonlindefs}).

Using the above definitions it is straightforward to write down the equations for $j=1/2$. These are
\begin{equation}
\left\lbrace\hat{\mathcal{H}}+\frac{7}{2}m+\frac{3}{4}T+
\left(\begin{array}{c|c}
\mathbf{A}+\mathbf{B} & \mathbf{C}\\
\hline
\mathbf{C}^{\dagger} &\mathbf{A}-\mathbf{B}
\end{array}\right)
\right\rbrace\begin{pmatrix}L_1^{-\tfrac{1}{2}}\\\vdots\\L_6^{-\tfrac{1}{2}}\\L_1^{\tfrac{1}{2}}\\\vdots\\L_6^{\tfrac{1}{2}}\end{pmatrix}=\mathcal{E}_m\begin{pmatrix}L_1^{-\tfrac{1}{2}}\\\vdots\\L_6^{-\tfrac{1}{2}}\\L_1^{\tfrac{1}{2}}\\\vdots\\L_6^{\tfrac{1}{2}}\end{pmatrix}
\label{eq:r1j12SE}
\end{equation}
with
\begin{equation}
\mathbf{B}\equiv \frac{1}{4}\begin{pmatrix}
\mathbf{y}_3\,\boldsymbol{\sigma}^3 & -2i\,\mathbf{z}_3\,\mathbb{1} & 0\\
2i\,\mathbf{z}_3\,\mathbb{1} &~~~\mathbf{y}_3\,\boldsymbol{\sigma}^3 &0\\
0 &0 &\mathbf{y}_3\,\boldsymbol{\sigma}^3
\end{pmatrix}
\end{equation}
and
\begin{equation}
\mathbf{C}\equiv\frac{1}{4}\begin{pmatrix}
\mathbf{y}_1\,\boldsymbol{\sigma}^1-i\,\mathbf{y}_2\,\boldsymbol{\sigma}^2 & 0& 2\,\mathbf{z}_2\,\mathbb{1} \\
0 &\mathbf{y}_1\,\boldsymbol{\sigma}^1-i\,\mathbf{y}_2\,\boldsymbol{\sigma}^2 &-2i\,\mathbf{z}_1\,\mathbb{1}\\
-2\,\mathbf{z}_2\,\mathbb{1} &2i\,\mathbf{z}_1\,\mathbb{1}&\mathbf{y}_1\,\boldsymbol{\sigma}^1-i\,\mathbf{y}_2\,\boldsymbol{\sigma}^2
\end{pmatrix}~.
\end{equation}

\subsection{R=2}
As we can see, the number of equations keeps increasing with fermion number and spin. Therefore in this section and the next, we will only give the reduced Schr\"{o}dinger equations for $j=0$. As before the general highest weight $R=2$ wavefunction admits a decomposition:
\begin{equation}
|\psi\rangle_2=e^{i\,j\,\vartheta_1}\sin^j\vartheta_2\sum_{p=-j}^j e^{i\,p\,\vartheta_3}\cot^p\left(\frac{\vartheta_2}{2}\right)\,f^{p}_{AB\alpha\beta}\,\bar{\chi}_A^\alpha\bar{\chi}_B^\beta|0\rangle~.
\end{equation}
In order to avoid over-counting let us set $f^p_{AB\alpha\beta}=0$ whenever $B<A$ and similarly $f^p_{AA\alpha\beta}=0$ (no sum on indices) whenever $\beta\leq\alpha$. Imposing that $J^3|\psi\rangle_2=j|\psi\rangle_2$, $J^+|\psi\rangle_2=0$ and that each term in the sum have definite $P^3$ eigenvalue imposes that the functions $f^p_{AB\alpha\beta}$ take on a particular form. These are (no sum on indices and $A<B$):
\begin{align}
f_{AA12}^p&=L^p_{A}\left(\mathbf{x}_a\right)~,\\
f^p_{AB12}&=\frac{e^{-i\vartheta_3}}{2}\sin\vartheta_2\,Y^p_{AB}\left(\mathbf{x}_a\right)+\cos^2\left(\frac{\vartheta_2}{2}\right) \,R_{AB}^p\left(\mathbf{x}_a\right)-\sin^2\left(\frac{\vartheta_2}{2}\right) \,S_{AB}^p\left(\mathbf{x}_a\right)-\frac{e^{i\vartheta_3}}{2}\sin\vartheta_2\, U^p_{AB}\left(\mathbf{x}_a\right)\\
f^p_{AB21}&=\frac{e^{-i\vartheta_3}}{2}\sin\vartheta_2\,Y^p_{AB}\left(\mathbf{x}_a\right)-\sin^2\left(\frac{\vartheta_2}{2}\right) \,R_{AB}^p\left(\mathbf{x}_a\right)+\cos^2\left(\frac{\vartheta_2}{2}\right) \,S_{AB}^p\left(\mathbf{x}_a\right)-\frac{e^{i\vartheta_3}}{2}\sin\vartheta_2\, U^p_{AB}\left(\mathbf{x}_a\right)\\
f^p_{AB11}&=e^{-i\vartheta_1}\left\lbrace e^{-i\vartheta_3}\cos^2\left(\frac{\vartheta_2}{2}\right)Y^p_{AB}\left(\mathbf{x}_a\right)-\frac{1}{2}\sin\vartheta_2\left(R_{AB}^p\left(\mathbf{x}_a\right)+S_{AB}^p\left(\mathbf{x}_a\right)\right)+e^{i\vartheta_3}\sin^2\left(\frac{\vartheta_2}{2}\right)U^p_{AB}\left(\mathbf{x}_a\right)\right\rbrace\\
f^p_{AB22}&=~\,e^{i\vartheta_1}\left\lbrace e^{-i\vartheta_3}\sin^2\left(\frac{\vartheta_2}{2}\right)Y^p_{AB}\left(\mathbf{x}_a\right)+\frac{1}{2}\sin\vartheta_2\left(R_{AB}^p\left(\mathbf{x}_a\right)+S_{AB}^p\left(\mathbf{x}_a\right)\right)+e^{i\vartheta_3}\cos^2\left(\frac{\vartheta_2}{2}\right)U^p_{AB}\left(\mathbf{x}_a\right)\right\rbrace .
\end{align}
Notice that even for $j=0$, determining the spectrum will involve solving a set of 15 coupled partial differential equations. We will label the set of functions $Y_{AB}^p\equiv Y_{6-A-B}^p$ and so on for the remaining functions. We also define the following vector of functions:
\begin{equation}
\Psi^0_{R=2}\equiv\left(L_1^0,\dots,R_1^0,\dots,S_1^0,\dots,U_1^0,\dots,Y_1^0,\dots\right)^\text{T}~.
\end{equation}
The $j=0$ Schr\"{o}dinger equation is then:
\begin{equation}
\left\lbrace\hat{\mathcal{H}}+\frac{5}{2}m+\frac{3}{4}T+\mathbf{D}+\mathbf{L}+g\mathbf{M}\right\rbrace\Psi^0_{R=2}
=\mathcal{E}_m\,\Psi^0_{R=2}
\label{eq:r2j0SE}
\end{equation}
where $\mathbf{D}$, $\mathbf{L}$ and $\mathbf{M}$ are $15\times 15$ matrices that can be written in terms of $3\times 3 $ blocks as follows
\begin{equation}
\mathbf{D}\equiv
\begin{pmatrix}
\mathbf{d}^1&0&0&0&0\\
0&\mathbf{d}^3&\mathbf{d}^1-\frac{\mathbf{y}_3}{4}\mathbb{1}&0&0\\
0&\mathbf{d}^1-\frac{\mathbf{y}_3}{4}\mathbb{1}&\mathbf{d}^3&0&0\\
0&0&0&-\mathbf{d}^3&\frac{1}{4}\left(\mathbf{y}_1-\mathbf{y}_2\right)\mathbb{1}\\
0&0&0&\frac{1}{4}\left(\mathbf{y}_1-\mathbf{y}_2\right)\mathbb{1}&-\mathbf{d}^3
\end{pmatrix}
\end{equation}
\begin{equation}
\mathbf{L}\equiv-\frac{1}{2}
\begin{pmatrix}
2\sum_a\mathbf{y}_a\,|\boldsymbol{\mathcal{L}}^a| &0 &0 &0 & 0\\
0& 0 &0 &\mathbf{z}_1\,\boldsymbol{\mathcal{L}}^1+i\,\mathbf{z}_2\,\boldsymbol{\mathcal{L}}^2&\mathbf{z}_1\,\boldsymbol{\mathcal{L}}^1-i\,\mathbf{z}_2\,\boldsymbol{\mathcal{L}}^2\\
0 & 0 &0  &\mathbf{z}_1\,\boldsymbol{\mathcal{L}}^1+i\,\mathbf{z}_2\,\boldsymbol{\mathcal{L}}^2 &\mathbf{z}_1\,\boldsymbol{\mathcal{L}}^1-i\,\mathbf{z}_2\,\boldsymbol{\mathcal{L}}^2\\
0& \mathbf{z}_1\,\boldsymbol{\mathcal{L}}^1-i\,\mathbf{z}_2\,\boldsymbol{\mathcal{L}}^2 &\mathbf{z}_1\,\boldsymbol{\mathcal{L}}^1-i\,\mathbf{z}_2\,\boldsymbol{\mathcal{L}}^2 &-2\,\mathbf{z}_3\,\boldsymbol{\mathcal{L}}^3 &0\\
0 & \mathbf{z}_1\,\boldsymbol{\mathcal{L}}^1+i\,\mathbf{z}_2\,\boldsymbol{\mathcal{L}}^2 &\mathbf{z}_1\,\boldsymbol{\mathcal{L}}^1+i\,\mathbf{z}_2\,\boldsymbol{\mathcal{L}}^2 &0 &2\,\mathbf{z}_3\,\boldsymbol{\mathcal{L}}^3\\
\end{pmatrix}
\end{equation}
\begin{equation}
\mathbf{M}\equiv\begin{pmatrix}
0 & \mathbf{x}_3\, \mathbf{m}^3 & \mathbf{x}_3\, \mathbf{m}^3 &\mathbf{x}_1\, \mathbf{m}^1+\mathbf{x}_2\, \mathbf{m}^2 & \mathbf{x}_2\, \mathbf{m}^2-\mathbf{x}_1\, \mathbf{m}^1\\
\mathbf{x}_3\,\mathbf{m}^{3\dagger} & -\mathbf{x}_3\,\boldsymbol{\mathcal{L}}^3 &0 &\mathbf{x}_2\,\mathbf{d}^2 &\mathbf{x}_2\,\mathbf{d}^{2\dagger}-\mathbf{x}_1\,\boldsymbol{\mathcal{L}}^1\\
\mathbf{x}_3\,\mathbf{m}^{3\dagger} & 0 &\mathbf{x}_3\,\boldsymbol{\mathcal{L}}^3 &-\mathbf{x}_1\,\boldsymbol{\mathcal{L}}^1-\mathbf{x}_2\,\mathbf{d}^{2\dagger} &-\mathbf{x}_2\,\mathbf{d}^{2}\\
\mathbf{x}_1\, \mathbf{m}^{1\dagger}+\mathbf{x}_2\, \mathbf{m}^{2\dagger} & \mathbf{x}_2\,\mathbf{d}^{2\dagger} &-\mathbf{x}_1\,\boldsymbol{\mathcal{L}}^1-\mathbf{x}_2\,\mathbf{d}^{2} &\mathbf{x}_3\,\boldsymbol{\mathcal{L}}^3 &0\\
\mathbf{x}_2\, \mathbf{m}^{2\dagger}-\mathbf{x}_1\, \mathbf{m}^{1\dagger} &\mathbf{x}_2\,\mathbf{d}^{2} -\mathbf{x}_1\,\boldsymbol{\mathcal{L}}^1 &-\mathbf{x}_2\,\mathbf{d}^{2\dagger} &0 &-\mathbf{x}_3\,\boldsymbol{\mathcal{L}}^3\\
\end{pmatrix}~.
\end{equation}
In these definitions, the $\boldsymbol{\mathcal{L}}^i$ are the $3\times3$ generators of $SO(3)$ defined below (\ref{polardecomp}). The $\mathbf{d}^i$ are
\begin{equation}
\mathbf{d}^1\equiv \left(\frac{T}{4}-\mathbf{y}_a\right)\delta_{ab}~,\quad\quad \mathbf{d}^2\equiv \begin{pmatrix}0 &0 &0\\
0 &0 &0 \\
-1 &0 &0\end{pmatrix}~,\quad\quad
\mathbf{d}^3\equiv\frac{1}{2}\left(\mathbf{y}_a-\frac{1}{2}\mathbf{y}_3\right)\delta_{ab}~,
\end{equation}
and the $\mathbf{m}^i$ are
\begin{equation}
\mathbf{m}^1\equiv \begin{pmatrix}
0 &0 &0\\
i &0 &0 \\
i &0 &0
\end{pmatrix}~,\quad\quad \mathbf{m}^2\equiv
\begin{pmatrix}
0 &-1 &0\\
0 &0 &0 \\
0 &-1 &0
\end{pmatrix}~,\quad\quad
\mathbf{m}^3\equiv\begin{pmatrix}
0 &0 &i\\
0 &0 &i \\
0 &0 &0
\end{pmatrix}~.
\end{equation}
Whenever a matrix appears in an absolute value symbol $|\cdot|$, the absolute value is to be applied to the entries of the matrix.

The Schr\"{o}dinger operator for $j=1/2$ will be a generalization of the above operator to one acting on 30 functions. We do not provide expressions for it here, but analyze its spectrum in the main text.

\subsection{R=3}
The highest weight $R=3$ wavefunctions take the form
\begin{equation}
|\psi\rangle_3=e^{i\,j\,\vartheta_1}\sin^j\vartheta_2\sum_{p=-j}^j e^{i\,p\,\vartheta_3}\cot^p\left(\frac{\vartheta_2}{2}\right)\,f^{p}_{ABC\alpha\beta\gamma}\,\bar{\chi}_A^\alpha\,\bar{\chi}_B^\beta\,\bar{\chi}_C^\gamma|0\rangle~.
\end{equation}
To avoid overcounting we set
\begin{align}
f^p_{ABC\alpha\beta\gamma}&=0\quad\quad\text{if }C<B \text{ or } B<A\\
f^p_{AAB\alpha\beta\gamma}&=0\quad\quad\text{if }\beta\leq\alpha\\
f^p_{ABB\alpha\beta\gamma}&=0\quad\quad\text{if }\gamma\leq\beta~.
\end{align}
Because of the fermionic statistics $f^p_{AAA\alpha\beta\gamma}=0$ identically. Imposing the highest weight condition forces $f^p_{123\alpha\beta\gamma}$ to take the following form
\begin{equation}
f^p_{123\alpha\beta\gamma}=\sum_{\substack{a,b, c=1}}^2 F^p_{abc}\left(\mathbf{x}_a\right)u_{\alpha a}\left(\vec{\vartheta}\right)u_{\beta b}\left(\vec{\vartheta}\right)u_{\gamma c}\left(\vec{\vartheta}\right)~,
\end{equation}
with
\begin{equation}
u_{\alpha a}\left(\vec{\vartheta}\right)\equiv e^{\frac{i}{2}\left((-1)^\alpha\vartheta_1+(-1)^a\vartheta_3\right)}\left\lbrace\left(1-|\alpha-a|\right)\cos\left(\frac{\vartheta_2}{2}\right)+(\alpha-a)\sin\left(\frac{\vartheta_2}{2}\right)\right\rbrace~.
\end{equation}
Furthermore
\begin{align}
f_{AAB\alpha\beta\gamma}^p &=U^p_{AAB}\left(\mathbf{x}_a\right)y^1_{\alpha\beta\gamma}\left(\vec{\vartheta}\right)+Y^p_{AAB}\left(\mathbf{x}_a\right)y^2_{\alpha\beta\gamma}\left(\vec{\vartheta}\right)\\
f_{ABB\alpha\beta\gamma}^p &=U^p_{ABB}\left(\mathbf{x}_a\right)y^1_{\alpha\beta\gamma}\left(\vec{\vartheta}\right)+Y^p_{ABB}\left(\mathbf{x}_a\right)y^2_{\alpha\beta\gamma}\left(\vec{\vartheta}\right)
\end{align}
where
\begin{align}
y^1_{\alpha\beta\gamma}\left(\vec{\vartheta}\right)&\equiv\frac{e^{\frac{i}{2}\left\lbrace\left((-1)^\alpha+(-1)^\beta+(-1)^\gamma\right)\vartheta_1-\vartheta_3\right\rbrace}}{2}\left[\left(4-\alpha\,\beta\,\gamma\right)\cos\left(\frac{\vartheta_2}{2}\right)+\left(\alpha\,\beta\,\gamma-2\right)\sin\left(\frac{\vartheta_2}{2}\right)\right]\\
y^2_{\alpha\beta\gamma}\left(\vec{\vartheta}\right)&\equiv\frac{e^{\frac{i}{2}\left\lbrace\left((-1)^\alpha+(-1)^\beta+(-1)^\gamma\right)\vartheta_1+\vartheta_3\right\rbrace}}{2}\left[\left(\alpha\,\beta\,\gamma-4\right)\sin\left(\frac{\vartheta_2}{2}\right)+\left(\alpha\,\beta\,\gamma-2\right)\cos\left(\frac{\vartheta_2}{2}\right)\right]~.
\end{align}
Notice that for $j=0$ the reduced Schr\"{o}dinger equation is a set of 20 coupled partial differential equations. We will give the Schr\"{o}dinger operator acting on the following vector of functions
\begin{multline}
\Psi_{R=3}^0\equiv\left(F^0_{122},F^0_{211},F^0_{121},F^0_{212},F^0_{221},F^0_{112},F^0_{111},F^0_{222},\right.\\\left. U^0_{113},U^0_{223},Y^0_{113},Y^0_{223},U_{112}^0,U^0_{233},Y^0_{112},Y^0_{233},U^0_{122},U^0_{133}Y^0_{122},Y^0_{133}\right)^\text{T}~.
\end{multline}
With $\Psi_{R=3}^0$ defined, we are tasked with solving the following set of differential equations:
\begin{equation}
\left\lbrace\hat{\mathcal{H}}+\frac{3}{2}m+\frac{1}{4}\left(\frac{11}{2}T-\mathbf{y}_3\right)-\mathbf{I}+\mathbf{J}+\mathbf{J}^\dagger+\mathbf{K}\right\rbrace\Psi_{R=3}^0=\mathcal{E}_m\,\Psi_{R=3}^0
\label{eq:r3j0SE}~,
\end{equation}
where $\mathbf{I}$, $\mathbf{J}$, and $\mathbf{K}$ are $20\times 20$ matrices that can be written in block form as follows:
\begin{equation}
\mathbf{I}\equiv
\begin{pmatrix}
\diagentry{\mathbf{y}_1\mathbb{1}_{2\times 2}}\\
&\diagentry{\mathbf{y}_2\mathbb{1}_{2\times 2}}&&&&0&\\
&&\diagentry{\mathbf{y}_3\mathbb{1}_{2\times 2}}\\
&&&\diagentry{\left(\mathbf{y}_1+\mathbf{y}_2\right)\mathbb{1}_{2\times 2}}\\
&&&&\diagentry{\frac{3}{4}\left(\mathbf{y}_1+\mathbf{y}_2\right)\mathbb{1}_{4\times 4}}\\
&0&&&&\diagentry{\frac{1}{2}\left(T+\frac{\mathbf{y}_1-\mathbf{y}_2}{2}\right)\mathbb{1}_{4\times 4}}\\
&&&&&&\diagentry{\frac{1}{2}\left(T-\frac{\mathbf{y}_1-\mathbf{y}_2}{2}\right)\mathbb{1}_{4\times 4}}~~~~~~~~
\end{pmatrix}~,
\end{equation}
and $\mathbf{J}$ and $\mathbf{K}$ can be written in terms of $4\times4$ blocks as follows:
\begin{equation}
\mathbf{J}\equiv
\frac{i}{2}\begin{pmatrix}
0&0&\mathbf{z}_3\,\mathbf{a}^3+2g\,\mathbf{x}_3\, \mathbf{b}^3 &~~~\mathbf{z}_2\,\mathbf{a}^2+2g\,\mathbf{x}_2\, \mathbf{b}^2 &\mathbf{z}_1\,\mathbf{a}^1+2g\,\mathbf{x}_1\, \mathbf{b}^1\\
0&0&0& -\mathbf{z}_2\,\mathbf{a}^2+2g\,\mathbf{x}_2\, \mathbf{c}^2 &\mathbf{z}_1\,\mathbf{e}^1+2g\,\mathbf{x}_1\, \mathbf{c}^1\\
0&0 &0 &\boldsymbol{\sigma}^1\otimes\left(\mathbf{z}_1\,\mathbb{1}+2g\,\mathbf{x}_1\, \boldsymbol{\sigma}^3\right) &-\boldsymbol{\sigma}^2\otimes\left(\mathbf{z}_2\,\boldsymbol{\sigma}^1-2i\,g\,\mathbf{x}_2\, \boldsymbol{\sigma}^2\right)\\
 0&0 &0 &0 &\boldsymbol{\sigma}^3\otimes\left(\mathbf{z}_3\,\mathbb{1}-2g\,\mathbf{x}_3\, \boldsymbol{\sigma}^3\right)\\
 0&0 &0 &0 &0
\end{pmatrix}~,
\end{equation}
\begin{equation}
\mathbf{K}\equiv
\begin{pmatrix}
\frac{1}{4}\left(T-5\,\mathbf{y}_3\right)\boldsymbol{\sigma}^1\otimes\boldsymbol{\sigma}^1 &\frac{1}{4}\left(\mathbf{y}_1\,\mathbf{s}+\mathbf{y}_2\,\mathbf{t}\right)&0&0&0\\
\frac{1}{4}\left(\mathbf{y}_1\,\mathbf{s}+\mathbf{y}_2\,\mathbf{t}\right)^\dagger&\frac{1}{4}\left(\mathbf{y}_1-\mathbf{y}_2\right)\boldsymbol{\sigma}^1\otimes\mathbb{1}&0&0&0\\
0&0&-\mathbf{y}_3~\mathbb{1}\otimes\boldsymbol{\sigma}^1 &0&0\\
0&0&0&-\mathbf{y}_2~\mathbb{1}\otimes\boldsymbol{\sigma}^1&0\\
0&0&0&0&-\mathbf{y}_1~\mathbb{1}\otimes\boldsymbol{\sigma}^1\\
\end{pmatrix}~,
\end{equation}
where we have implicitly defined
\begin{align*}
&\mathbf{a}^1\equiv -\frac{1}{2}\begin{pmatrix} 0& 0\\1 &1\end{pmatrix}\otimes\left(\mathbb{1}-\boldsymbol{\sigma}^1\right)+\frac{1}{2}\begin{pmatrix} 0& 0\\1 &-1\end{pmatrix}\otimes\left(-i\,\boldsymbol{\sigma}^2+\boldsymbol{\sigma}^3\right)~,&
&\mathbf{b}^1\equiv \frac{1}{2}\begin{pmatrix} -1& 1\\0 &0\end{pmatrix}\otimes\left(\mathbb{1}+\boldsymbol{\sigma}^1\right)-\frac{1}{2}\begin{pmatrix} 1& 1\\0 &0\end{pmatrix}\otimes\left(i\,\boldsymbol{\sigma}^2+\boldsymbol{\sigma}^3\right)~,& \\
&\mathbf{a}^2\equiv -\frac{i}{2}\begin{pmatrix} -1& 1\\0 &0\end{pmatrix}\otimes\left(\mathbb{1}-\boldsymbol{\sigma}^1\right)+\frac{i}{2}\begin{pmatrix} 1& 1\\0 &0\end{pmatrix}\otimes\left(-i\,\boldsymbol{\sigma}^2+\boldsymbol{\sigma}^3\right)~,&
&\mathbf{b}^2\equiv -\frac{i}{2}\begin{pmatrix} 0& 0\\1 &1\end{pmatrix}\otimes\left(\mathbb{1}+\boldsymbol{\sigma}^1\right)+\frac{i}{2}\begin{pmatrix} 0& 0\\1 &-1\end{pmatrix}\otimes\left(i\,\boldsymbol{\sigma}^2+\boldsymbol{\sigma}^3\right)~,& \\
&\mathbf{a}^3\equiv i\,\boldsymbol{\sigma}^2\otimes\begin{pmatrix}-1 &1\\ 0 & 0\end{pmatrix}+\boldsymbol{\sigma}^3\otimes\begin{pmatrix} 0 & 0\\-1 &1\end{pmatrix}~,&
&\mathbf{b}^3\equiv -\mathbb{1}\otimes\begin{pmatrix} 0 & 0\\1 &1\end{pmatrix}-\boldsymbol{\sigma}^1\otimes\begin{pmatrix}1 &1\\ 0 & 0\end{pmatrix}~,&
\end{align*}
as well as
\begin{align*}
&\mathbf{c}^1\equiv \frac{1}{2}\begin{pmatrix} 0& 0\\1 &-1\end{pmatrix}\otimes\left(\mathbb{1}+\boldsymbol{\sigma}^1\right)+\frac{1}{2}\begin{pmatrix} 0& 0\\1 &1\end{pmatrix}\otimes\left(i\,\boldsymbol{\sigma}^2+\boldsymbol{\sigma}^3\right)~, \\
&\mathbf{c}^2\equiv -\frac{i}{2}\begin{pmatrix} 0& 0\\1 &1\end{pmatrix}\otimes\left(\mathbb{1}+\boldsymbol{\sigma}^1\right)-\frac{i}{2}\begin{pmatrix} 0& 0\\1 &-1\end{pmatrix}\otimes\left(i\,\boldsymbol{\sigma}^2+\boldsymbol{\sigma}^3\right)~, \\
&\mathbf{e}^1\equiv -\frac{1}{2}\begin{pmatrix} 1& 1\\0 &0\end{pmatrix}\otimes\left(\mathbb{1}-\boldsymbol{\sigma}^1\right)+\frac{1}{2}\begin{pmatrix} -1& 1\\0 &0\end{pmatrix}\otimes\left(-i\,\boldsymbol{\sigma}^2+\boldsymbol{\sigma}^3\right)~,
\end{align*}
and finally
\begin{equation*}
\mathbf{s}\equiv \begin{pmatrix}1 &1\\ 0&0\end{pmatrix}\otimes\mathbb{1}+\begin{pmatrix}0&0\\ -3 &1\end{pmatrix}\otimes\boldsymbol{\sigma}^1\quad\quad\quad\text{and}\quad\quad\quad
\mathbf{t}\equiv \begin{pmatrix}-3 &1\\ 0 &0\end{pmatrix}\otimes\mathbb{1}+\begin{pmatrix}0 & 0\\ 1 &-1\end{pmatrix}\otimes\boldsymbol{\sigma}^1~.
\end{equation*}
The Hamiltonian acting on the $R=3$, $j=1/2$ wavefunction will be a generalization of the above operator to one acting on 40 functions. We will not give the expression here, but we analyze the spectrum of the $R=3$, $j=1/2$ sector numerically in the main text.

\section{Metric on the moduli space}\label{ap:metric}
In order to determine the one-loop effective action for the $\nu=0$ theory, we follow \cite{Douglas:1996yp,Becker:1997wh,Taylor:1998tv} and pass to the Lagrangian formulation of our gauge-quantum mechanics, including gauge-fixing terms and ghosts. We will use the background field method \cite{Abbott:1981ke,Abbott:1980hw}---that is we will expand the fields $X^i_A=B^i_A+\tilde{X}^i_A$ where $B^i_A$ is a fixed background field configuration and $\tilde{X}^i_A$ are the fluctuating degree of freedom. We choose $B^i_A=\delta_{A3}\,\vec{\mathbf{x}}$ such that it parametrizes motion along the moduli space.

The gauge-fixed Lagrangian is:
\begin{equation}
  \mathcal{L}=\mathcal{L}_{\rm bos.}+\mathcal{L}_{\rm ferm.}+\mathcal{L}_{\rm g.f.}+\mathcal{L}_{\rm ghost}
\end{equation}
with
\begin{align}
\mathcal{L}_{\rm bos.}&=\frac{1}{2}\left(\mathcal{D}_t X^i_A\right)^2-\frac{g^2}{4}\left(f_{ABC}\,X^i_B\,X^j_C\right)^2\\
\mathcal{L}_{\rm ferm.}&=i\left(\bar{\lambda}_A\,\mathcal{D}_t\lambda_A-g\,f_{ABC}\bar{\lambda}_A\,X^k_B\,\boldsymbol{\sigma}^k\lambda_C\right)\\
\mathcal{L}_{\rm g.f.}&=-\frac{1}{2\xi}\left(\mathcal{D}_t^{\rm bg} A_A\right)^2\\
\mathcal{L}_{\rm ghost}&=\bar{c}_A\left(-\delta_{AB}\,\partial_t^2-g\,f_{ACB}\,\partial_t\left(A_C\cdot\right)+g^2\,f_{ACD}f_{DEB}B^i_C\,X^i_E\right)c_B
\end{align}
and
\begin{equation}
\mathcal{D}_t X^i_A\equiv\dot{X}^i_A+g\,f_{ABC}\,A_B\,X^i_C~,\quad\quad
\mathcal{D}_t \lambda_{A\alpha}\equiv\dot{\lambda}_{A\alpha}+g\,f_{ABC}\,A_B\,\lambda_{C\alpha}~,\quad\quad
\mathcal{D}_t^{\rm bg} A_A\equiv-\dot{A}_A+g\,f_{ABC}\,B^i_B\,X^i_C~.
\end{equation}
 We further set $\xi=1$, corresponding to Feynman gauge.
 We can obtain the correction to the metric on moduli space by choosing a background field $\vec{\mathbf{x}}$ as follows \cite{Douglas:1996yp}
\begin{equation}
\vec{\mathbf{x}}=(b,vt,0)
\end{equation}
where $b$ is to be thought of as an impact parameter for a particle moving at speed $v$. We now Wick rotate $t\rightarrow -i\tau$, $v\rightarrow i\gamma$ and $A_A\rightarrow i A_A$ and expand the action to quadratic order in fluctuating fields about the background field $B^i_A=\delta_{A3}\,\vec{\mathbf{x}}$. The idea is to integrate out all fields that obtain a mass, through interaction with the background field, at one loop.

Following this procedure, it is easy to show that all fields with color index $A=3$ remain massless, while the rest obtain time dependent masses. After diagonalizing the mass-matrix for the bosonic fields we find the contribution to the Euclidean effective action coming respectively from the bosonic, fermionic and ghost determinants are:
\begin{align}
\delta S_E^{\rm bos.}=&-2\, \text{Tr}\,\log\left(-\partial_\tau^2+g^2\left(b^2+\gamma^2\tau^2\right)\right)-\, \text{Tr}\,\log\left(-\partial_\tau^2+g^2\left(b^2+\gamma^2\tau^2\right)-2g\,\gamma\right)
\nonumber\\&~~~~~~~~~~~~~~~~~~~~~~~~~~~~~~~~~~~~~~~~~~~~~~~~~~~~~~~~~~~~~~~~~~~~-\, \text{Tr}\,\log\left(-\partial_\tau^2+g^2\left(b^2+\gamma^2\tau^2\right)+2g\,\gamma\right)\\
\delta S_E^{\rm ferm.}=&~\text{Tr}\,\log\begin{pmatrix}\partial_\tau &-g(\gamma\,\tau+ib)\\ -g(\gamma\,\tau-ib) &\partial_\tau\end{pmatrix}+\text{Tr}\,\log\begin{pmatrix}\partial_\tau &g(\gamma\,\tau-ib)\\ g(\gamma\,\tau+ib) &\partial_\tau\end{pmatrix}~\\
\delta S_E^{\rm ghost}=&~2\, \text{Tr}\,\log\left(-\partial_\tau^2+g^2\left(b^2+\gamma^2\tau^2\right)\right)~.
\end{align}
Note that the ghost determinant cancels against the contribution coming from four of the eight massive bosons. Up to a diverging constant, which will cancel between the bosonic and fermionic terms, we can replace $\log(\lambda)=-\int_0^\infty \frac{ds}{s}e^{-s\,\lambda}$ and, summing over the spectra of the above differential operators, we find
\begin{align}
\delta S_E&=\int_0^\infty \frac{ds}{s} e^{-b^2g^2s}\left(\cosh(2g \gamma s)\,\text{csch}(g \gamma s)-\coth(g \gamma s)\right)\\
&=\int_0^\infty \frac{ds}{s} e^{-b^2g^2s}\text{sech}\left(\frac{g \gamma s}{2}\right)\,\sinh\left(\frac{3g \gamma s}{2}\right)~.
\end{align}
Let us now wick rotate back to Lorentzian time and use:
\begin{equation}
\frac{e^{-b^2g^2s}}{s}=\int \frac{dt}{\sqrt{\pi\,s}}g\,v\,e^{-s g^2 r^2}~,\quad\quad\quad r^2={b^2+v^2t^2}~,
\end{equation}
to write down the Lorentzian action to $O\left(v^2\right)$:
\begin{align}
  i\,S_L&=i\int dt\left[ \frac{v^2}{2}-g\,v\,\int \frac{ds}{\sqrt{\pi\,s}}e^{-s g^2 r^2}\text{sec}\left(\frac{g v s}{2}\right)\,\sin\left(\frac{3g v s}{2}\right)\right]\\
 &=i\int dt~\frac{1}{2}\left(1-\frac{3}{2\,g\,r^3}\right)v^2+O\left(v^4\right)~,
\end{align}
which is the same correction as found in \cite{Smilga:2002mra,Akhmedov:2002mi}. It also resembles the correction to the moduli space metric in the D0-D4 system \cite{Douglas:1996yp}, albeit with a different coefficient and sign.

\bibliographystyle{utphys}
\bibliography{Matrix_QM}{}

\providecommand{\href}[2]{#2}\begingroup\raggedright\begin{thebibliography}{10}

\bibitem{Claudson:1984th}
M.~Claudson and M.~B. Halpern, ``{Supersymmetric Ground State Wave
  Functions},''
\href{http://dx.doi.org/10.1016/0550-3213(85)90500-0}{{\em Nucl. Phys.}
  {\bfseries B250} (1985) 689--715}.

\bibitem{Denef:2002ru}
F.~Denef, ``{Quantum quivers and Hall / hole halos},''
  \href{http://dx.doi.org/10.1088/1126-6708/2002/10/023}{{\em JHEP} {\bfseries
  10} (2002) 023},
\href{http://arxiv.org/abs/hep-th/0206072}{{\ttfamily arXiv:hep-th/0206072
  [hep-th]}}.

\bibitem{Asplund:2015yda}
C.~T. Asplund, F.~Denef, and E.~Dzienkowski, ``{Massive quiver matrix models
  for massive charged particles in AdS},''
  \href{http://dx.doi.org/10.1007/JHEP01(2016)055}{{\em JHEP} {\bfseries 01}
  (2016) 055},
\href{http://arxiv.org/abs/1510.04398}{{\ttfamily arXiv:1510.04398 [hep-th]}}.

\bibitem{Banks:1996vh}
T.~Banks, W.~Fischler, S.~H. Shenker, and L.~Susskind, ``{M theory as a matrix
  model: A Conjecture},''
  \href{http://dx.doi.org/10.1103/PhysRevD.55.5112}{{\em Phys. Rev.} {\bfseries
  D55} (1997) 5112--5128},
\href{http://arxiv.org/abs/hep-th/9610043}{{\ttfamily arXiv:hep-th/9610043
  [hep-th]}}.

\bibitem{Berenstein:2002jq}
D.~E. Berenstein, J.~M. Maldacena, and H.~S. Nastase, ``{Strings in flat space
  and pp waves from N=4 superYang-Mills},''
  \href{http://dx.doi.org/10.1088/1126-6708/2002/04/013}{{\em JHEP} {\bfseries
  04} (2002) 013},
\href{http://arxiv.org/abs/hep-th/0202021}{{\ttfamily arXiv:hep-th/0202021
  [hep-th]}}.

\bibitem{Sethi:1997pa}
S.~Sethi and M.~Stern, ``{D-brane bound states redux},''
  \href{http://dx.doi.org/10.1007/s002200050374}{{\em Commun. Math. Phys.}
  {\bfseries 194} (1998) 675--705},
\href{http://arxiv.org/abs/hep-th/9705046}{{\ttfamily arXiv:hep-th/9705046
  [hep-th]}}.

\bibitem{Moore:1998et}
G.~W. Moore, N.~Nekrasov, and S.~Shatashvili, ``{D particle bound states and
  generalized instantons},''
  \href{http://dx.doi.org/10.1007/s002200050016}{{\em Commun. Math. Phys.}
  {\bfseries 209} (2000) 77--95},
\href{http://arxiv.org/abs/hep-th/9803265}{{\ttfamily arXiv:hep-th/9803265
  [hep-th]}}.

\bibitem{Yi:1997eg}
P.~Yi, ``{Witten index and threshold bound states of D-branes},''
  \href{http://dx.doi.org/10.1016/S0550-3213(97)00486-0}{{\em Nucl. Phys.}
  {\bfseries B505} (1997) 307--318},
\href{http://arxiv.org/abs/hep-th/9704098}{{\ttfamily arXiv:hep-th/9704098
  [hep-th]}}.

\bibitem{Lee:2016dbm}
S.-J. Lee and P.~Yi, ``{Witten Index for Noncompact Dynamics},''
  \href{http://dx.doi.org/10.1007/JHEP06(2016)089}{{\em JHEP} {\bfseries 06}
  (2016) 089},
\href{http://arxiv.org/abs/1602.03530}{{\ttfamily arXiv:1602.03530 [hep-th]}}.

\bibitem{deWit:1997ab}
B.~de~Wit, ``{Supersymmetric quantum mechanics, supermembranes and Dirichlet
  particles},'' \href{http://dx.doi.org/10.1016/S0920-5632(97)00312-5}{{\em
  Nucl. Phys. Proc. Suppl.} {\bfseries 56B} (1997) 76--87},
\href{http://arxiv.org/abs/hep-th/9701169}{{\ttfamily arXiv:hep-th/9701169
  [hep-th]}}.

\bibitem{Polchinski:1999br}
J.~Polchinski, ``{M theory and the light cone},''
  \href{http://dx.doi.org/10.1143/PTPS.134.158}{{\em Prog. Theor. Phys. Suppl.}
  {\bfseries 134} (1999) 158--170},
\href{http://arxiv.org/abs/hep-th/9903165}{{\ttfamily arXiv:hep-th/9903165
  [hep-th]}}.

\bibitem{Anous:2015xah}
T.~Anous, ``{SUSY in Silico: numerical D-brane bound state spectroscopy},''
\href{http://arxiv.org/abs/1511.01450}{{\ttfamily arXiv:1511.01450 [hep-th]}}.

\bibitem{Wosiek:2002nm}
J.~Wosiek, ``{Spectra of supersymmetric Yang-Mills quantum mechanics},''
  \href{http://dx.doi.org/10.1016/S0550-3213(02)00810-6}{{\em Nucl. Phys.}
  {\bfseries B644} (2002) 85--112},
\href{http://arxiv.org/abs/hep-th/0203116}{{\ttfamily arXiv:hep-th/0203116
  [hep-th]}}.

\bibitem{Campostrini:2004bs}
M.~Campostrini and J.~Wosiek, ``{High precision study of the structure of D=4
  supersymmetric Yang-Mills quantum mechanics},''
  \href{http://dx.doi.org/10.1016/j.nuclphysb.2004.10.022}{{\em Nucl. Phys.}
  {\bfseries B703} (2004) 454--498},
\href{http://arxiv.org/abs/hep-th/0407021}{{\ttfamily arXiv:hep-th/0407021
  [hep-th]}}.

\bibitem{Anninos:2013mfa}
D.~Anninos, T.~Anous, F.~Denef, and L.~Peeters, ``{Holographic
  Vitrification},'' \href{http://dx.doi.org/10.1007/JHEP04(2015)027}{{\em JHEP}
  {\bfseries 04} (2015) 027},
\href{http://arxiv.org/abs/1309.0146}{{\ttfamily arXiv:1309.0146 [hep-th]}}.

\bibitem{Smilga:1984jg}
A.~V. Smilga, ``{WITTEN INDEX CALCULATION IN SUPERSYMMETRIC GAUGE THEORY},''
  \href{http://dx.doi.org/10.1016/0550-3213(86)90176-8}{{\em Nucl. Phys.}
  {\bfseries B266} (1986) 45--57}.
[Yad. Fiz.42,728(1985)].

\bibitem{Goldstone:1978he}
J.~Goldstone and R.~Jackiw, ``{Unconstrained Temporal Gauge for Yang-Mills
  Theory},''
\href{http://dx.doi.org/10.1016/0370-2693(78)90065-5}{{\em Phys. Lett.}
  {\bfseries B74} (1978) 81--84}.

\bibitem{Klebanov:1991qa}
I.~R. Klebanov, ``{String theory in two-dimensions},'' in {\em {Spring School
  on String Theory and Quantum Gravity (to be followed by Workshop) Trieste,
  Italy, April 15-23, 1991}}, pp.~30--101.
\newblock 1991.
\newblock
\href{http://arxiv.org/abs/hep-th/9108019}{{\ttfamily arXiv:hep-th/9108019
  [hep-th]}}.
\newblock

\bibitem{Halpern:1997fv}
M.~B. Halpern and C.~Schwartz, ``{Asymptotic search for ground states of SU(2)
  matrix theory},'' \href{http://dx.doi.org/10.1142/S0217751X98002110}{{\em
  Int. J. Mod. Phys.} {\bfseries A13} (1998) 4367--4408},
\href{http://arxiv.org/abs/hep-th/9712133}{{\ttfamily arXiv:hep-th/9712133
  [hep-th]}}.

\bibitem{Smilga:1989ew}
A.~V. Smilga, ``{Super-Yang-Mills quantum mechanics and supermembrane
  spectrum},'' in {\em {Trieste Conference on Supermembranes and Physics in 2+1
  Dimensions Trieste, Italy, July 17-21, 1989}}, pp.~0182--195.
\newblock 1989.
\newblock \href{http://arxiv.org/abs/1406.5987}{{\ttfamily arXiv:1406.5987
  [hep-th]}}.
\newblock
\url{http://inspirehep.net/record/1302441/files/arXiv:1406.5987.pdf}.
\newblock

\bibitem{Smilga:2002mra}
A.~V. Smilga, ``{Born-Oppenheimer corrections to the effective zero mode
  Hamiltonian in SYM theory},''
  \href{http://dx.doi.org/10.1088/1126-6708/2002/04/054}{{\em JHEP} {\bfseries
  04} (2002) 054},
\href{http://arxiv.org/abs/hep-th/0201048}{{\ttfamily arXiv:hep-th/0201048
  [hep-th]}}.

\bibitem{Akhmedov:2002mi}
E.~T. Akhmedov and A.~V. Smilga, ``{On the relation between effective
  supersymmetric actions in different dimensions},''
  \href{http://dx.doi.org/10.1134/1.1634333}{{\em Phys. Atom. Nucl.} {\bfseries
  66} (2003) 2238--2244}, \href{http://arxiv.org/abs/hep-th/0202027}{{\ttfamily
  arXiv:hep-th/0202027 [hep-th]}}.
[Yad. Fiz.66,2290(2003)].

\bibitem{Frohlich:1999zf}
J.~Frohlich, G.~M. Graf, D.~Hasler, J.~Hoppe, and S.-T. Yau, ``{Asymptotic form
  of zero energy wave functions in supersymmetric matrix models},''
  \href{http://dx.doi.org/10.1016/S0550-3213(99)00649-5}{{\em Nucl. Phys.}
  {\bfseries B567} (2000) 231--248},
\href{http://arxiv.org/abs/hep-th/9904182}{{\ttfamily arXiv:hep-th/9904182
  [hep-th]}}.

\bibitem{Graf:1998bm}
G.~M. Graf and J.~Hoppe, ``{Asymptotic ground state for 10-dimensional reduced
  supersymmetric SU(2) Yang-Mills theory},''
\href{http://arxiv.org/abs/hep-th/9805080}{{\ttfamily arXiv:hep-th/9805080
  [hep-th]}}.

\bibitem{Ivanov:1990jn}
E.~A. Ivanov and A.~V. Smilga, ``{Supersymmetric gauge quantum mechanics:
  Superfield description},''
\href{http://dx.doi.org/10.1016/0370-2693(91)90862-K}{{\em Phys. Lett.}
  {\bfseries B257} (1991) 79--82}.

\bibitem{Diaconescu:1997ut}
D.-E. Diaconescu and R.~Entin, ``{A Nonrenormalization theorem for the d = 1,
  N=8 vector multiplet},''
  \href{http://dx.doi.org/10.1103/PhysRevD.56.8045}{{\em Phys. Rev.} {\bfseries
  D56} (1997) 8045--8052},
\href{http://arxiv.org/abs/hep-th/9706059}{{\ttfamily arXiv:hep-th/9706059
  [hep-th]}}.

\bibitem{Douglas:1996yp}
M.~R. Douglas, D.~N. Kabat, P.~Pouliot, and S.~H. Shenker, ``{D-branes and
  short distances in string theory},''
  \href{http://dx.doi.org/10.1016/S0550-3213(96)00619-0}{{\em Nucl. Phys.}
  {\bfseries B485} (1997) 85--127},
\href{http://arxiv.org/abs/hep-th/9608024}{{\ttfamily arXiv:hep-th/9608024
  [hep-th]}}.

\bibitem{Anninos:2013nra}
D.~Anninos, T.~Anous, P.~de~Lange, and G.~Konstantinidis, ``{Conformal quivers
  and melting molecules},''
  \href{http://dx.doi.org/10.1007/JHEP03(2015)066}{{\em JHEP} {\bfseries 03}
  (2015) 066},
\href{http://arxiv.org/abs/1310.7929}{{\ttfamily arXiv:1310.7929 [hep-th]}}.

\bibitem{Lin:2014wka}
Y.-H. Lin and X.~Yin, ``{On the Ground State Wave Function of Matrix Theory},''
  \href{http://dx.doi.org/10.1007/JHEP11(2015)027}{{\em JHEP} {\bfseries 11}
  (2015) 027},
\href{http://arxiv.org/abs/1402.0055}{{\ttfamily arXiv:1402.0055 [hep-th]}}.

\bibitem{Taylor:2001vb}
W.~Taylor, ``{M(atrix) theory: Matrix quantum mechanics as a fundamental
  theory},'' \href{http://dx.doi.org/10.1103/RevModPhys.73.419}{{\em Rev. Mod.
  Phys.} {\bfseries 73} (2001) 419--462},
\href{http://arxiv.org/abs/hep-th/0101126}{{\ttfamily arXiv:hep-th/0101126
  [hep-th]}}.

\bibitem{Susskind:1998vk}
L.~Susskind, ``{Holography in the flat space limit},''
  \href{http://arxiv.org/abs/hep-th/9901079}{{\ttfamily arXiv:hep-th/9901079
  [hep-th]}}.
[AIP Conf. Proc.493,98(1999)].

\bibitem{Balasubramanian:1997kd}
V.~Balasubramanian, R.~Gopakumar, and F.~Larsen, ``{Gauge theory, geometry and
  the large N limit},''
  \href{http://dx.doi.org/10.1016/S0550-3213(98)00377-0}{{\em Nucl. Phys.}
  {\bfseries B526} (1998) 415--431},
\href{http://arxiv.org/abs/hep-th/9712077}{{\ttfamily arXiv:hep-th/9712077
  [hep-th]}}.

\bibitem{Anninos:2016szt}
D.~Anninos, T.~Anous, and F.~Denef, ``{Disordered Quivers and Cold Horizons},''
  \href{http://dx.doi.org/10.1007/JHEP12(2016)071}{{\em JHEP} {\bfseries 12}
  (2016) 071},
\href{http://arxiv.org/abs/1603.00453}{{\ttfamily arXiv:1603.00453 [hep-th]}}.

\bibitem{KitaevTalks}
A.~Kitaev, ``{A simple model of quantum holography},'' {\em KITP strings
  seminar and Entanglement program (Feb. 12, April 7, and May 27,)} (2015) .
  \url{http://online.kitp.ucsb.edu/online/entangled15/}.

\bibitem{Polchinski:2016xgd}
J.~Polchinski and V.~Rosenhaus, ``{The Spectrum in the Sachdev-Ye-Kitaev
  Model},'' \href{http://dx.doi.org/10.1007/JHEP04(2016)001}{{\em JHEP}
  {\bfseries 04} (2016) 001},
\href{http://arxiv.org/abs/1601.06768}{{\ttfamily arXiv:1601.06768 [hep-th]}}.

\bibitem{Maldacena:2016hyu}
J.~Maldacena and D.~Stanford, ``{Remarks on the Sachdev-Ye-Kitaev model},''
  \href{http://dx.doi.org/10.1103/PhysRevD.94.106002}{{\em Phys. Rev.}
  {\bfseries D94} no.~10, (2016) 106002},
\href{http://arxiv.org/abs/1604.07818}{{\ttfamily arXiv:1604.07818 [hep-th]}}.

\bibitem{Jensen:2016pah}
K.~Jensen, ``{Chaos in AdS$_2$ Holography},''
  \href{http://dx.doi.org/10.1103/PhysRevLett.117.111601}{{\em Phys. Rev.
  Lett.} {\bfseries 117} no.~11, (2016) 111601},
\href{http://arxiv.org/abs/1605.06098}{{\ttfamily arXiv:1605.06098 [hep-th]}}.

\bibitem{Sachdev:2015efa}
S.~Sachdev, ``{Bekenstein-Hawking Entropy and Strange Metals},''
  \href{http://dx.doi.org/10.1103/PhysRevX.5.041025}{{\em Phys. Rev.}
  {\bfseries X5} no.~4, (2015) 041025},
\href{http://arxiv.org/abs/1506.05111}{{\ttfamily arXiv:1506.05111 [hep-th]}}.

\bibitem{Jevicki:2016bwu}
A.~Jevicki, K.~Suzuki, and J.~Yoon, ``{Bi-Local Holography in the SYK Model},''
  \href{http://dx.doi.org/10.1007/JHEP07(2016)007}{{\em JHEP} {\bfseries 07}
  (2016) 007},
\href{http://arxiv.org/abs/1603.06246}{{\ttfamily arXiv:1603.06246 [hep-th]}}.

\bibitem{Cotler:2016fpe}
J.~S. Cotler, G.~Gur-Ari, M.~Hanada, J.~Polchinski, P.~Saad, S.~H. Shenker,
  D.~Stanford, A.~Streicher, and M.~Tezuka, ``{Black Holes and Random
  Matrices},''
\href{http://arxiv.org/abs/1611.04650}{{\ttfamily arXiv:1611.04650 [hep-th]}}.

\bibitem{Witten:2016iux}
E.~Witten, ``{An SYK-Like Model Without Disorder},''
\href{http://arxiv.org/abs/1610.09758}{{\ttfamily arXiv:1610.09758 [hep-th]}}.

\bibitem{Gurau:2016lzk}
R.~Gurau, ``{The complete $1/N$ expansion of a SYK--like tensor model},''
\href{http://arxiv.org/abs/1611.04032}{{\ttfamily arXiv:1611.04032 [hep-th]}}.

\bibitem{Klebanov:2016xxf}
I.~R. Klebanov and G.~Tarnopolsky, ``{Uncolored Random Tensors, Melon Diagrams,
  and the SYK Models},''
\href{http://arxiv.org/abs/1611.08915}{{\ttfamily arXiv:1611.08915 [hep-th]}}.

\bibitem{Krishnan:2016bvg}
C.~Krishnan, S.~Sanyal, and P.~N. Bala~Subramanian, ``{Quantum Chaos and
  Holographic Tensor Models},''
\href{http://arxiv.org/abs/1612.06330}{{\ttfamily arXiv:1612.06330 [hep-th]}}.

\bibitem{Maldacena:2015waa}
J.~Maldacena, S.~H. Shenker, and D.~Stanford, ``{A bound on chaos},''
  \href{http://dx.doi.org/10.1007/JHEP08(2016)106}{{\em JHEP} {\bfseries 08}
  (2016) 106},
\href{http://arxiv.org/abs/1503.01409}{{\ttfamily arXiv:1503.01409 [hep-th]}}.

\bibitem{Anninos:2016klf}
D.~Anninos and G.~A. Silva, ``{Solvable Quantum Grassmann Matrices},''
\href{http://arxiv.org/abs/1612.03795}{{\ttfamily arXiv:1612.03795 [hep-th]}}.

\bibitem{Ferrari:2017ryl}
F.~Ferrari, ``{The Large D Limit of Planar Diagrams},''
\href{http://arxiv.org/abs/1701.01171}{{\ttfamily arXiv:1701.01171 [hep-th]}}.

\bibitem{Gur-Ari:2015rcq}
G.~Gur-Ari, M.~Hanada, and S.~H. Shenker, ``{Chaos in Classical D0-Brane
  Mechanics},'' \href{http://dx.doi.org/10.1007/JHEP02(2016)091}{{\em JHEP}
  {\bfseries 02} (2016) 091},
\href{http://arxiv.org/abs/1512.00019}{{\ttfamily arXiv:1512.00019 [hep-th]}}.

\bibitem{Balthazar:2016utu}
B.~Balthazar, V.~A. Rodriguez, and X.~Yin, ``{Hamiltonian Truncation Study of
  Supersymmetric Quantum Mechanics: S-Matrix and Metastable States},''
\href{http://arxiv.org/abs/1610.07275}{{\ttfamily arXiv:1610.07275 [hep-th]}}.

\bibitem{Ferrell:1987gf}
R.~C. Ferrell and D.~M. Eardley, ``{Slow motion scattering and coalescence of
  maximally charged black holes},''
\href{http://dx.doi.org/10.1103/PhysRevLett.59.1617}{{\em Phys. Rev. Lett.}
  {\bfseries 59} (1987) 1617}.

\bibitem{Michelson:1999zf}
J.~Michelson and A.~Strominger, ``{The Geometry of (super)conformal quantum
  mechanics},'' \href{http://dx.doi.org/10.1007/PL00005528}{{\em Commun. Math.
  Phys.} {\bfseries 213} (2000) 1--17},
\href{http://arxiv.org/abs/hep-th/9907191}{{\ttfamily arXiv:hep-th/9907191
  [hep-th]}}.

\bibitem{Michelson:1999dx}
J.~Michelson and A.~Strominger, ``{Superconformal multiblack hole quantum
  mechanics},'' \href{http://dx.doi.org/10.1088/1126-6708/1999/09/005}{{\em
  JHEP} {\bfseries 09} (1999) 005},
\href{http://arxiv.org/abs/hep-th/9908044}{{\ttfamily arXiv:hep-th/9908044
  [hep-th]}}.

\bibitem{BrittoPacumio:1999ax}
R.~Britto-Pacumio, J.~Michelson, A.~Strominger, and A.~Volovich, ``{Lectures on
  Superconformal Quantum Mechanics and Multi-Black Hole Moduli Spaces},''
  \href{http://dx.doi.org/10.1007/978-94-011-4303-5_6}{{\em NATO Sci. Ser. C}
  {\bfseries 556} (2000) 255--284},
\href{http://arxiv.org/abs/hep-th/9911066}{{\ttfamily arXiv:hep-th/9911066
  [hep-th]}}.

\bibitem{Asatrian:1982ga}
G.~M. Asatryan and G.~K. Savvidy, ``{Configuration Manifold of {Yang-Mills}
  Classical Mechanics},''
\href{http://dx.doi.org/10.1016/0375-9601(83)90887-3}{{\em Phys. Lett.}
  {\bfseries A99} (1983) 290}.

\bibitem{Savvidy:1982wx}
G.~K. Savvidy, ``{YANG-MILLS CLASSICAL MECHANICS AS A KOLMOGOROV K SYSTEM},''
\href{http://dx.doi.org/10.1016/0370-2693(83)91146-2}{{\em Phys. Lett.}
  {\bfseries B130} (1983) 303--307}.

\bibitem{Savvidy:1984gi}
G.~K. Savvidy, ``{YANG-MILLS QUANTUM MECHANICS},''
\href{http://dx.doi.org/10.1016/0370-2693(85)90260-6}{{\em Phys. Lett.}
  {\bfseries B159} (1985) 325--329}.

\bibitem{Becker:1997wh}
K.~Becker and M.~Becker, ``{A Two loop test of M(atrix) theory},''
  \href{http://dx.doi.org/10.1016/S0550-3213(97)00518-X}{{\em Nucl. Phys.}
  {\bfseries B506} (1997) 48--60},
\href{http://arxiv.org/abs/hep-th/9705091}{{\ttfamily arXiv:hep-th/9705091
  [hep-th]}}.

\bibitem{Taylor:1998tv}
W.~Taylor and M.~Van~Raamsdonk, ``{Supergravity currents and linearized
  interactions for matrix theory configurations with fermionic backgrounds},''
  \href{http://dx.doi.org/10.1088/1126-6708/1999/04/013}{{\em JHEP} {\bfseries
  04} (1999) 013},
\href{http://arxiv.org/abs/hep-th/9812239}{{\ttfamily arXiv:hep-th/9812239
  [hep-th]}}.

\bibitem{Abbott:1981ke}
L.~F. Abbott, ``{Introduction to the Background Field Method},''
{\em Acta Phys. Polon.} {\bfseries B13} (1982) 33.

\bibitem{Abbott:1980hw}
L.~F. Abbott, ``{The Background Field Method Beyond One Loop},''
\href{http://dx.doi.org/10.1016/0550-3213(81)90371-0}{{\em Nucl. Phys.}
  {\bfseries B185} (1981) 189--203}.

\end{thebibliography}\endgroup

\end{document}